\documentclass{article}
\usepackage{graphicx}

\usepackage{multirow}
\usepackage{tikz}
\usepackage{algorithm}
\usepackage{algorithmic}
\usepackage{comment}
\usepackage[usenames,dvipsnames]{pstricks}
\usepackage{epsfig}

\newcommand{\cale}{\ensuremath{{\mathcal{E}}}}
\newcommand{\electionsystem}{\cale}

\newcommand{\strictlybigger}{\ensuremath{\mathit{StrictlyBigger}}}
\newcommand{\bigger}{\ensuremath{\mathit{Bigger}}}
\newcommand{\netadv}{\ensuremath{\mathit{D}}}
\usepackage{ifthen}
\usepackage{latexsym}
\usepackage{amsfonts}
\usepackage{amsmath}
\usepackage{amssymb}
\usepackage{pifont}

\usetikzlibrary{arrows}

\newcommand{\hugeDebug}{false}
\ifthenelse{\equal{\hugeDebug}{false}}{
\newcommand{\normalspacing}{\singlespacing}
\pagestyle{plain}
\setlength{\oddsidemargin}{0.25in}
\setlength{\evensidemargin}{\oddsidemargin}
\setlength{\textwidth}{6in}
\setlength{\textheight}{8in}
\setlength{\topmargin}{-0.0in}
}
{
\newcommand{\normalspacing}{\niceninespacing}
\pagestyle{headings}
 \setlength{\oddsidemargin}{-0.4in}
 \setlength{\evensidemargin}{\oddsidemargin}
 \setlength{\textwidth}{5.3in}
 \setlength{\textheight}{6in}
 \setlength{\textheight}{6.25in}
 \setlength{\topmargin}{-0.0in}
}

\newlength{\filength}
\newsavebox{\gcbox}
\sbox{\gcbox}{\framebox[\filength]{\rule{0ex}{2ex}}}

  \newtheorem{theorem}{Theorem}[section]
  \newtheorem{corollary}[theorem]{Corollary}

\newcommand\qedblob{\ding{113}}
\def\literalqed{{\ \nolinebreak\hfill\mbox{\qedblob\quad}}}

  \newtheorem{definition}[theorem]{Definition}

\showhyphens{}

\newcount\hour  \newcount\minutes  \hour=\time  \divide\hour by 60
\minutes=\hour  \multiply\minutes by -60  \advance\minutes by \time
\def\mmmddyyyy{\ifcase\month\or Jan\or Feb\or Mar\or Apr\or May\or Jun\or Jul\or
  Aug\or Sep\or Oct\or Nov\or Dec\fi \space\number\day, \number\year}
\def\hhmm{\ifnum\hour<10 0\fi\number\hour :%
  \ifnum\minutes<10 0\fi\number\minutes}

\makeatletter
\def\@citex[#1]#2{\if@filesw\immediate\write\@auxout{\string\citation{#2}}\fi
  \def\@citea{}\@cite{\@for\@citeb:=#2\do
    {\@citea\def\@citea{,\linebreak[0]}\@ifundefined
       {b@\@citeb}{{\bf ?}\@warning
       {Citation `\@citeb' on page \thepage \space undefined}}%
\hbox{\csname b@\@citeb\endcsname}}}{#1}}
\newcommand{\singlespacing}{\let\CS=
\@currsize\renewcommand{\baselinestretch}{1}\tiny\CS}
\newcommand{\singlespacingplus}{\let\CS=
\@currsize\renewcommand{\baselinestretch}{1.25}\tiny\CS}
\newcommand{\doublespacing}{\let\CS=
\@currsize\renewcommand{\baselinestretch}{1.75}\tiny\CS}
\newcommand{\extradoublespacing}{\let\CS=
\@currsize\renewcommand{\baselinestretch}{1.9}\tiny\CS}
\newcommand{\nicenicespacing}{\let\CS=
\@currsize\renewcommand{\baselinestretch}{1.9}\tiny\CS}
\newcommand{\draftspacing}{\let\CS=
\@currsize\renewcommand{\baselinestretch}{2.0}\tiny\CS}
\newcommand{\hugedraftspacing}{\let\CS=
\@currsize\renewcommand{\baselinestretch}{2.4}\tiny\CS}
\newcommand{\niceonespacing}{\let\CS=\@currsize\renewcommand{\baselinestretch}{1.1}\tiny\CS}
\newcommand{\nicetwospacing}{\let\CS=\@currsize\renewcommand{\baselinestretch}{1.2}\tiny\CS}
\newcommand{\nicethreespacing}{\let\CS=\@currsize\renewcommand{\baselinestretch}{1.3}\tiny\CS}
\newcommand{\singlespacingplusplus}{\let\CS=\@currsize\renewcommand{\baselinestretch}{1.35}\tiny\CS}
\newcommand{\nicefourspacing}{\let\CS=\@currsize\renewcommand{\baselinestretch}{1.4}\tiny\CS}
\newcommand{\nicefivespacing}{\let\CS=\@currsize\renewcommand{\baselinestretch}{1.5}\tiny\CS}
\newcommand{\nicesixspacing}{\let\CS=\@currsize\renewcommand{\baselinestretch}{1.6}\tiny\CS}
\newcommand{\nicesevenspacing}{\let\CS=\@currsize\renewcommand{\baselinestretch}{1.7}\tiny\CS}
\newcommand{\niceeightspacing}{\let\CS=\@currsize\renewcommand{\baselinestretch}{1.8}\tiny\CS}
\newcommand{\niceninespacing}{\let\CS=\@currsize\renewcommand{\baselinestretch}{1.9}\tiny\CS}
\makeatother%

\normalspacing



\newcommand{\p}{{\rm P}}

\newcommand{\np}{{\rm NP}}

\newenvironment{proofs}{\noindent{\it Proof.}\hspace*{0.5em}}{\literalqed\bigskip}
\newenvironment{proofsof}[1]{\noindent{\it Proof {#1}.}\hspace*{0.5em}}{\literalqed\bigskip}

\newcommand{\sectionref}[1]{Section~\ref{#1}}

\DeclareMathSymbol{\subsetneq}{\mathbin}{AMSb}{"28}
\DeclareMathSymbol{\supsetneq}{\mathbin}{AMSb}{"29}

\newcommand{\pisnp}{\ensuremath{\p=\np}}

\newcommand{\bigo}{{\protect\cal O}}

\newcommand{\pref}{\mathit{pref}}

\begin{document}
\sloppy

\title{Schulze and Ranked-Pairs Voting Are Fixed-Parameter Tractable to
Bribe, Manipulate, and Control\thanks{A two-page extended-abstract 
version of this paper
appeared in 
the Proceedings of the 12th International Conference on Autonomous Agents and Multiagent Systems (AAMAS~2013)~\cite{hem-lav-men:c:schulze-and-ranked-pairs}.}}

\author{Lane A. Hemaspaandra,
Rahman Lavaee\\
Department of Computer Science \\
University of Rochester\\
Rochester,~NY~14627, USA
\and
Curtis Menton\thanks{Work done in part 
while at 
the 
University of Rochester's 
Department of Computer Science.}\\
Google Inc.\\
Mountain View, CA 94043}

\date{October 25, 2012; revised 
June 21, 2014}

\maketitle

\begin{abstract}
  Schulze and ranked-pairs elections have received much attention recently,
and the former has quickly 
become a
quite 
widely used 
election system.
For many cases
  these systems have been proven resistant to 
  bribery, control, 
or manipulation,
  with ranked pairs being particularly praised for being
  NP-hard for all three of those.  
Nonetheless, the present paper shows that with respect to the 
number of candidates, 
Schulze and 
ranked-pairs elections are fixed-parameter tractable 
to bribe, control, and manipulate: we obtain uniform,
polynomial-time algorithms whose degree does not depend on the number
of candidates.
We also provide such algorithms for 
some weighted variants of these problems.
\end{abstract}

\section{Introduction}
Schulze voting~\cite{sch:j:schulze}, though relatively recently proposed, has
quickly been rather widely adopted.
Designed in part to well-handle
candidate cloning, its users include the Wikimedia foundation, the
Pirate Party in a dozen countries, Debian, KDE, the Free Software
Foundation Europe, and dozens of other 
organizations, and Wikipedia 
even asserts that 
``currently the Schulze method is the most widespread Condorcet 
method''~\cite{wik:url:schulze-voting}.

Although the winner-choosing process in Schulze voting is a bit
complicated to describe, involving minima, maxima, and comparisons of
paths in the so-called weighted majority graph,
Schulze~\cite{sch:j:schulze} proved 
that 
finding who won a Schulze election 
nonetheless is polynomial-time computable,
and 
Parkes and Xia~\cite{par-xia:c:ranked-pairs} 
for the so-called destructive case and 
Gaspers et al.~\cite{gas-kal-nar-wal:c:schulze} 
for the so-called constructive case (extending 
a one-manipulator result for that case by
Parkes and Xia~\cite{par-xia:c:ranked-pairs}) proved 
that the (unweighted coalitional) manipulation problem
for Schulze elections is 
polynomial-time 
computable.
On the other hand, 
Parkes and Xia~\cite{par-xia:c:ranked-pairs} 
proved that 
for Schulze elections bribery is
NP-hard, 
and the work of 
Parkes and Xia~\cite{par-xia:c:ranked-pairs} 
and 
Menton and 
Singh~\cite{men-sin:c:schulze} 
established that for Schulze elections 15 of the 22 
benchmark
control attacks are NP-hard.

Parkes and Xia also note that, by the work of~\cite{par-xia:c:ranked-pairs,con-pro-ros-xia-zuc:c:unweighted-manipulation,xia:margin-of-victory},
the ranked-pairs election system, which is not widely popular but like
Schulze has a polynomial-time winner-determination problem and like
Schulze is based on the weighted majority graph, 
is resistant to
(basically, NP-hard with respect to) bribery, control under each of the
control types they study in their paper, and manipulation.  Based on
their discovery 
that ranked pairs is more broadly resistant to attacks than 
Schulze, the fact that Schulze itself ``is in wide use,'' and the
fact that there is ``broad axiomatic support for both Schulze and
ranked pairs,'' Parkes and Xia~\cite{par-xia:c:ranked-pairs}
quite reasonably conclude that ``there
seems to be good support to adopt ranked pairs in practical
applications.''

However, in this paper we show that the resistances-to-attack of Schulze
and ranked pairs are both quite fragile.  

For each of the
\emph{bribery} and 
\emph{control} cases
studied by Parkes and Xia,
Menton and Singh, and Gaspers et al.~for which 
they did not already prove Schulze voting 
to be in P,
we prove that Schulze voting is
fixed-parameter tractable with respect to the number of 
candidates.  
(The (unweighted)
\emph{manipulation} cases were already 
all put into P by these papers.)
Fixed-parameter tractable (see~\cite{nie:b:invitation-fpt})
means 
there is an algorithm for the problem whose running time is
$f(j)I^{O(1)}$, where $j$ is the number of candidates and $I$ is the
input's size.  This of course implies that for each fixed number of
candidates, the problems are in polynomial time, but it says much
more; it implies that there is a global bound 
on the degree of the polynomial running time,
regardless of what the fixed number of
candidates is.

That result might lead one to even more strongly suggest the adoption
of ranked pairs as an attractive alternative to Schulze.  However,
although for ranked pairs Parkes and Xia proved all the types of
bribery, control, and manipulation they studied to be NP-hard, we show
that every one of those cases is fixed-parameter tractable (with
respect to the number of candidates) for ranked pairs.  So even
ranked pairs does not offer a safe haven from fixed-parameter 
tractability.

Our final results section, Section~\ref{s:weighted}, looks
at bribery and manipulation in the case of 
\emph{weighted} voting, and proves a number of results for 
that case.  For example, for ranked pairs, we prove 
that weighted constructive coalitional manipulation is 
fixed-parameter tractable
with respect to the combined parameter
``number of candidates'' and ``cardinality of the 
manipulators' weight set.''  We give evidence that this 
fixed-parameter tractability 
result cannot be extended to a general P result, namely, we prove 
that 
weighted constructive coalitional manipulation is 
NP-complete for five or more candidates.
We also show that this ``five'' is optimal unless $\p=\np$, by 
proving that this problem is in polynomial
time for four or fewer candidates.

\section{Presentation of the Key Idea}\label{ss:key}
Our fixed-parameter 
tractability 
proofs are of interest in their own right, because they face a
very specific challenge, which at first might not even seem
possible to handle.
We now 
describe 
in 
relatively high-level terms 
what that challenge is and how we handle it.

Before we start that explanation, we need to present the 
definition of Schulze voting.
Voters will always vote by linear orders over the candidates; in doing
that, we adopt the complete, tie-free ordering case of Schulze used in
the papers most related to this
one~\cite{par-xia:c:ranked-pairs,men-sin:c:schulze,gas-kal-nar-wal:c:schulze}.
Given the
input set of candidates and the set of votes over them (as linear
orders), the weighted majority graph (WMG) is the graph that for each
ordered pair of candidates $c$ and $d$, $c\neq d$, has an edge from $c$
to $d$ having weight equal to the number of voters who prefer $c$ to
$d$ minus the number of voters who prefer $d$ to $c$.  
Clearly, either all WMG edges have even weight or all WMG edges have
odd weight, and the weight of the edge from $c$ to $d$ is negative one
times the weight of the edge from $d$ to $c$.  The ``strength'' of a
directed path between two nodes in the WMG is the
minimum weight of all the edges along that path.  The strength can be
negative.  The Schulze election system is that candidate $c$ is a
winner exactly if for each other candidate $d$ it holds that there is some
simple path from $c$ to $d$ whose strength is at least as great as
that of every simple path from $d$ to $c$.  A lovely result is that
the set of winners, under this definition, is always
nonempty~\cite{sch:j:schulze}.

We now give a small Schulze-election example 
from Parkes and Xia~\cite{par-xia:c:ranked-pairs}, over the candidate set 
$\{$\textit{1,2,3,4}$\}$. Although the votes are not specified here,
using 
McGarvey's method~\cite{mcg:j:election-graph}
we can build 
a profile of votes realizing the WMG of 
Figure~\ref{schulze-rp-example}.
We will carefully explain 
McGarvey's method in Appendix~\ref{sec:McGarvey}; 
for now we ask the reader to 
briefly take our claim on faith.
\begin{figure}[tp]
\begin{center}
\scalebox{1.0} %
{
\begin{pspicture}(0,-1.9410918)(5.406875,1.9410918)
\rput(1.565625,0.5889082){\textit{4}}
\rput(4.7704687,0.5889082){\textit{3}}
\rput(1.5704688,-1.6110919){\textit{1}}
\rput(4.7603126,-1.6110919){\textit{2}}
\psbezier[linewidth=0.04,arrowsize=0.05291667cm
  2.0,arrowlength=1.4,arrowinset=0.4]{->}(1.8825,-1.5210918)(2.4825,-1.3210918)(3.8825,-1.3210918)(4.4825,-1.5210918)
\psbezier[linewidth=0.04,arrowsize=0.05291667cm
  2.0,arrowlength=1.4,arrowinset=0.4]{->}(4.4825,0.47890818)(3.8825,0.2789082)(2.4825,0.2789082)(1.8825,0.47890818)
\psbezier[linewidth=0.04,arrowsize=0.05291667cm
  2.0,arrowlength=1.4,arrowinset=0.4]{->}(1.4825,0.2789082)(1.2825,-0.3210918)(1.2825,-0.7210918)(1.4825,-1.3210918)
\psbezier[linewidth=0.04,arrowsize=0.05291667cm
  2.0,arrowlength=1.4,arrowinset=0.4]{<-}(4.8825,0.2789082)(5.0825,-0.3210918)(5.0825,-0.7210918)(4.8825,-1.3210918)
\psbezier[linewidth=0.04,arrowsize=0.05291667cm
  2.0,arrowlength=1.4,arrowinset=0.4]{->}(1.8825,0.0789082)(2.0825,-0.5210918)(3.6825,-1.3210918)(4.4825,-1.3210918)
\rput(3.1659374,0.4889082){2}
\rput(1.173125,-0.091091804){6}
\rput(4.7025,-0.49109182){8}
\rput(3.1325,-1.6910918){8}
\rput(3.078125,-0.7110918){4}
\rput(0.973125,1.0089082){6}
\psbezier[linewidth=0.04,arrowsize=0.05291667cm
  2.0,arrowlength=1.4,arrowinset=0.4]{->}(4.7825,0.8789082)(4.4825,1.7789083)(1.3526967,1.9210918)(0.6825,1.1789082)(0.012303393,0.43672457)(0.35642397,-1.2437549)(1.2825,-1.6210918)
\psbezier[linewidth=0.04, arrowsize=0.05291667cm 2.0,arrowlength=1.4,arrowinset=0.4]{->}(1.5,-1.952521)(1.0790517,-2.33353)(4.535785,-2.3147144)(5.355017,-1.7412521)(6.1742487,-1.1677897)(5.95091,1.3682121)(4.955017,0.887479)
\rput[bl](5.855017,-1.7412521){0}
\end{pspicture} 
}

\caption{\label{schulze-rp-example}WMG for the election examples, with 
one edge from each pair left implicit. 
Those edges are the reverse edges of the displayed edges, with a weight equal to negative one times 
the weight of their displayed counterpart, e.g., the implicit edge from 
candidate 2 to candidate 4 has weight -4 and the implicit edge from
candidate 3 to candidate 1 has weight 0.}
\end{center}
\end{figure}
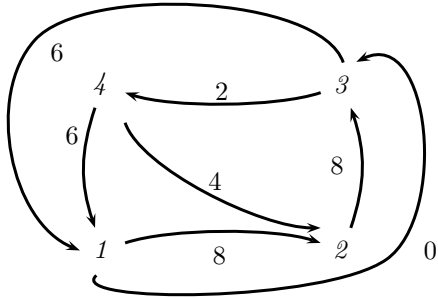
Candidate \textit{4} is the sole Schulze
winner, strictly beating each other candidate
in best-path strength.
For each other candidate $i$, 
candidate \textit{4} has 
a path to $i$ of strength $6$,
but $i$'s strongest path to $\textit{4}$ has strength $2$.

One of the most powerful tools to use in building algorithms
establishing fixed-parameter tractability is a result due to Lenstra,
showing that the integer linear programming feasibility problem 
(henceforth, ILPFP) is in P
if the number of variables is fixed~\cite{len:j:integer-fixed}.
Lenstra's result, based on the geometry of numbers, is very deep and
so 
strong that intuition whispers it should not even be
true; yet it is.

Now, if within 
an
appropriate-sized 
integer linear program with a
number of variables that was bounded by some function of the number of
candidates 
we could capture our bribery/manipulation/control
challenges and the action of the election system, we would be home
free.  Indeed, this has been done, for control, for such systems
as plurality, veto, Borda, Dodgson, and others (see the
discussion on p.~338 of~\cite{fal-hem-hem:j:multiprong}).
However, Schulze and ranked-pairs elections have such extremely
demanding definitions that they seem well beyond such an 
approach, and we have not been able to make that approach work.
So we have a challenge.

Fortunately, the literature provides a way to hope to approach even
systems that are too hard to directly wedge---together with the
manipulative action---into an ILPFP\@.  
That approach is to define some sort of structure associated
with subcases of behavior/outcomes of an election system, such that
for each fixed number of candidates the number of such structures is
bounded as a function of the number of candidates (independent of the
number of voters), yet such that for each such structure we can wedge
into an ILPFP the question
of whether the given action can be made to succeed in the system in a
way that is consistent with that structure.  If that can be done, then
we just loop over all such structures (for the given number of
candidates), and for each of them build and run the appropriate 
ILPFP.

This has not been done often, but it has been done for example
by Faliszewski et
al.~\cite{fal-hem-hem-rot:j:llull}, with respect to some control
problems, for the election system known as Copeland voting.  And the
structure they used is what they called a Copeland Output Table, which
is a collection of bits associated with the outcomes of the pairwise
majority contests between the candidates.

Unfortunately, such output tables don't seem to have 
enough information to support the case of Schulze or ranked pairs.
(However, see the comments at the start of 
Appendix~\ref{sec:alternate-swcf}.)
The natural structure that would allow us to tackle our systems is the
one the systems are based on, namely, the WMG, and
looping over all of those would allow us 
within the loop
to easily write/run 
an appropriate 
ILPFP
to check
the given case.  However, 
that
falls apart because the number of WMGs is
\emph{not} bounded as a function of the number of candidates; the
number also grows as a function of the number of voters.  
The impossibility
of looping over WMGs leaves us still faced with the challenge of how to 
tackle our problems.

A key 
contribution of this paper is to show that 
the needle described above can be threaded---and to 
thread it---for Schulze
and ranked-pairs elections.  In particular, we need to, for each of 
those election systems, find
a (winner-set certification) 
structure that on one hand is rich enough that for each 
structure instance we can within an 
ILPFP
check whether the given manipulative action can lead to
success in a way consistent with the case of which 
the particular instance of the
structure is speaking.  Yet on the other hand, the structure must
be so restrictive that the number of such structures is bounded purely as
a function of the number of candidates (independent of the number of
voters).  In brief, we need to find, if one exists, a ``sweet spot''
that meets both these competing needs.

We achieve this with structures we call Schulze winner-set
certification frameworks (SWCFs) and 
ranked pairs winner-set certification
frameworks (RPWCFs).
A 
Schulze winner-set certification framework
contains a
``pattern'' for how we can prove that a given set of candidates is
the winner set of a Schulze election.  
To do that, the structure for
each winner $a$ specifies, for each other candidate $b$, a ``strong path'' 
$\gamma_{ab}$ from $a$ to
$b$ in the WMG 
(recall that victory in Schulze elections is based on having
strong paths), 
and then---to establish that the other candidate $b$ has no stronger path
back to $a$---for every simple path from
$b$ back to our candidate $a$ the 
structure identifies a
``weak link'' (a directed edge on that path)
that will keep the path from being too strong;  to 
be more specific, we mean an edge on that path in the WMG 
such that its weight is less than or equal to that of 
every edge in our allegedly quite strong path $\gamma_{ab}$.
(Now, keep
in mind, at the time we are looping through the structure, we will not
even know how strong each link is, as the manipulation/bribery/control
will not yet even have happened; rather, the structure is specifying a
particular pattern of victory, and the 
ILPFP
will have to check whether the given type/amount
of manipulation/bribery/control can bring to life that victory
pattern.)  Additionally, 
for each candidate $a$ 
the structure claims is not a winner, the structure 
will specify what rival $b$ 
eliminates that candidate from the winner set and then
will outline a pattern for a proof that that is the case, in particular
giving a ``strong path'' from $b$ to $a$ and for each simple path
from $a$ to $b$ our structure will specify a ``weak link,'' i.e., an edge on
that path from $a$ to $b$ whose weight in the WMG we hope will be
\emph{strictly} less 
than the weight of all edges in the selected strong path from $b$ to $a$;
if all our hopes of this sort turn out to be true (and that is among what 
the integer linear program will be testing, for each of our
certification framework's structures), 
this proves that $b$ eliminates $a$.
Crucially, the number of 
structures in that Schulze winner-set certification framework,
though large, is bounded as
a function of the number of candidates. 
The certification framework, however, does not itself have its hands
on the weights of the WMG, and so the paths and edges it specifies are
all given in terms of the self-loop-free 
graph, on nodes named
$1,2,\ldots,\|C\|$, that between each pair of distinct 
nodes has edges in both directions.
Since the candidate names are irrelevant in 
Schulze voting, we can change to those canonical names, so that our
Schulze structures are always in terms of those 
names.

Crucially, as noted above the number of 
structures in our Schulze winner-set certification framework,
though large, is bounded as
a function of the number of candidates.  Yet, also crucially, this 
approach 
provides enough structure to allow a polynomial-sized (technically,
it is actually a uniform-over-all-numbers-of-candidates polynomial
multiplied by a constant that may depend on the number of candidates;
this does not trivialize the claim, since the ILPFP must work for 
all numbers of voters)
ILPFP
to do the ``rest'' of the work,
namely, to see whether by a given type of attack we can bring to life
the proof framework that a given instance of the structure sets out,
as to who the winners/nonwinners are in the Schulze election and why.
Appendix~\ref{sec:alternate-swcf} gives an alternate 
winner certification framework that readers may wish to look at to 
get more of a sense of what certification frameworks do and how they
may look.  That appendix's certification
framework is closer in flavor to Copeland
Output Tables than the approach above is, although as the appendix discusses
the alternative approach is both quite clear and quite subtle.

For ranked pairs, the entire approach is what we just described above,
except the certification framework we use is completely different than
that used for Schulze.  Ranked pairs is a method that is defined in
highly sequential terms, through successive rounds some of which add a
relationship between two candidates, and so our certification
framework will be making extensive guesses about what happens in each
round (and about a number of other things).  But again, we will ensure
that the number of such certification structures is bounded as a
function of the number of candidates (independent of the number of
voters), yet each structure will give 
enough information that the rest of the work can be
done by an integer linear programming feasibility problem.  Our notion of 
a 
ranked pairs winner-set certification
framework will be given in detail in Section~\ref{s:ranked-pairs}.

\section{Definitions}\label{s:defs}
Schulze elections were defined in the previous section.  We now
define the quite different system known as ranked pairs, due to 
Tideman
(see~\cite{tid:b:voting}).  
The ranked-pairs 
winner is defined by a sequential process that uses  
the weighted majority graph (WMG)\@.  
We choose the edge in the WMG of
greatest weight, say from $a$ to $b$, and fix in the eventual output
that $a$ must beat $b$; cases of ties, either 
regarding what edge has the greatest weight, or regarding cases 
where $(a,b)$ and $(b,a)$ both are weight zero, are handled 
as will be specified in 
footnote~\ref{f:ties}.  We then remove the edges between $a$ and $b$
from the WMG\@.  We then iterate this process, except if the greatest
remaining edge is one between two candidates who are already ordered
by earlier fixings of output ordering 
(this can happen, due to transitivity
applied to earlier fixings),
then we discard the pair of edges between those candidates.  We
continue until we have completely fixed a linear 
order.\footnote{\label{f:ties}There
  are two different types of ties that must be handled.  One is when
  we get to a case when we are considering an edge, and we don't
  discard it, and the candidates tie (the edges between them are both
  0); here, we break ties using some simple ordering among the
  candidates.  By simple, we mean feasible; there is a polynomial-time 
  machine
  that, given the candidates, outputs a linear ordering of them that
  is the ordering to use when breaking ties of this sort.  
  The second type of tie is when there is a tie as to what is the
  largest edge remaining in the WMG\@.  In ties of that sort, we
  use a simple---again, by simple we mean feasible, analogously to the
  first case---ordering among all unordered pairs of candidates to decide
  which pair having a highest-weight edge still left is the one to next
  consider.  
 If that pair is $\{a,b\}$ and both $(a,b)$ and $(b,a)$ 
   have weight zero, either edge can be chosen to consider next, 
   since which we consider at this point among $(a,b)$ and $(b,a)$ 
   makes no difference in the result of this step.

  An at first seemingly tempting alternate approach to breaking ties
  would be to
  require as part of the input the two types of tie-breaking orders
  discussed above.  But that is highly unattractive, since that would
  require changing the definitions of long-defined problems
  (manipulation, control, bribery), in order to add that extra 
  input part.  In truth, the tie-breaking is being made, by us and the earlier
  papers, to be a part of ranked pairs;  and so it should be a feature
  or setting that is part of one's version of ranked pairs, and should 
  not be built in by hacking the notions of manipulative actions.
  So to us, if one wants to speak about ranked pairs, one to
  be clear and complete must also specify the two feasible
  tie-breaking functions that are needed to completely define the
  system.  However, our main results for ranked pairs, which are 
  fixed-parameter tractability results, will all hold for all feasible
  tie-breaking functions.

  Here is an example of a pair of feasible tie-breaking functions.
  One could 
  break ties between two candidates in favor of the lexicographically
  larger.  And one could break ties between two candidate pairs 
  in favor 
  of the pair with the lexicographically-larger larger-candidate-of-the-pair,
  and when the larger members
  are the same in both pairs then breaking the tie in favor of 
  whichever pair has the lexicographically-larger 
  smaller-candidate-of-the-pair.  
  The suggestion to use the candidate-vs.-candidate 
  ordering to induce an ordering on the pairs---a suggestion our 
  example is consistent with---was made and used by
  the creator of ranked pairs, Nicolaus Tideman, 
  in his book ``Collective Decisions and Voting: The Potential
  for Public Choice''~\cite{tid:b:voting}.}
The candidate at the top of this linear order is the winner under 
ranked pairs.  Even if the first removed edge is from $a$ to $b$ and that 
edge has positive weight, it is possible that $a$ will not be the ranked 
pairs winner.

We give a small example of selecting the winner under ranked pairs.
We again consider the election with candidate set
$\{$\textit{1,2,3,4}$\}$ and votes such that
Figure~\ref{schulze-rp-example} is the WMG\@.  We will break 
order-of-consideration ties (due to tied edge 
weights) between $\{a,b\}$ and $\{c,d\}$ 
in favor of which pair 
has the lexicographically-larger 
larger-candidate-of-the-pair, and if they tie in that, on which has the 
lexicographically-larger smaller-candidate-of-the-pair.
Thus we handle the edges in the following order: 
\textit{3}$\rightarrow$\textit{2},
\textit{2}$\rightarrow$\textit{1}, \textit{4}$\rightarrow$\textit{1},
\textit{1}$\rightarrow$\textit{3}, \textit{4}$\rightarrow$\textit{2},
\textit{3}$\rightarrow$\textit{4}.  The output ordering
will be set by those (\textit{3}$\succ$\textit{2}, \textit{2}$\succ$\textit{1}, etc.),\ except 
with \textit{1}$\rightarrow$\textit{3} discarded due to
transitivity.  So under ranked pairs, 
\textit{3} is the sole winner.

As mentioned earlier, our elections are specified by a set of
candidates and voters (each vote is a tie-free linear ordering of the
candidates).  The standard (also called ``nonsuccinct'') approach to
the votes is that each comes in separately.  In the succinct approach,
which is meaningful only for systems such as Schulze and ranked
pairs that don't care about voters' names, each tie-free linear
ordering that is cast by at least one voter comes with, as a
binary integer, the number of voters that voted that way.  

In our problems we will speak of making a candidate $p$ a winner or
precluding $p$ from being a winner. This is
known as the nonunique-winner model or, in some papers,
the co-winner model.  If one changes ``a winner'' into
``the one and only winner,'' that is what is known as the
unique-winner model.

The problem definitions we are about to give present the 
definitions for the nonsuccinct, nonunique-winner case.  However,
the above two paragraphs 
make clear the very slight changes needed to 
define the 
succinct, nonunique-winner case, the 
nonsuccinct, unique-winner case, and the 
succinct, unique-winner case.  For consistency with the literature,
the wording of many of our attack-problem definitions is taken
directly from or modeled on the definitions
given in~\cite{fal-hem-hem-rot:j:llull}.  In these definitions,
$\cale$ will represent the election system.

\begin{description}
\item[Name:] The constructive bribery problem 
for $\cale$ elections
and the 
destructive bribery problem for $\cale$ 
elections~\cite{fal-hem-hem:j:bribery}.
\item[Given:] A set $C$ of candidates, a
collection
$V$ of voters 
(specified via their tie-free linear orders over $C$), 
a distinguished candidate $p \in C$, and a nonnegative integer $k$.
\item[Question (constructive):] 
Is it possible 
to change the votes of at most $k$ members of $V$ in such a way
that 
$p$ is a winner of the election 
whose voter collection is $V$ 
and whose candidate set is $C$.
\item[Question (destructive):] 
Is it possible 
to change the votes of at most $k$ members of $V$ in such a way
that 
$p$ is not a winner of the election 
whose voter collection is $V$ 
and whose candidate set is $C$.
\end{description}

In the literature, the above problem is often called
the unweighted, unpriced bribery problem.

\begin{description}
\item[Name:] The constructive manipulation problem 
for $\cale$ elections
and the 
destructive manipulation problem for $\cale$ 
elections~\cite{bar-tov-tri:j:manipulating,con-lan-san:j:when-hard-to-manipulate}.
\item[Given:] A set $C$ of candidates, a
collection $V$ of voters 
(specified via their tie-free linear orders over $C$),
a 
voter collection 
$W$ (each starting as a blank slate),
and a distinguished candidate $p \in C$.
\item[Question (constructive):] 
Is it possible 
to assign (tie-free linear order) votes to the members of $W$ in 
such a way that 
$p$ is a winner of the election whose 
voters are the members of $V$ 
and $W$, and whose candidate set is $C$.
\item[Question (destructive):] 
Is it possible 
to assign (tie-free linear order) votes to the members of $W$ in 
such a way that 
$p$ is not a winner of the election whose voters are the members of $V$ 
and $W$, and whose candidate set is $C$.
\end{description}

In the
literature, this is often 
called the unweighted coalitional manipulation problem.
We will sometimes refer to the members of the manipulative coalition $W$ 
as manipulators and to the members of $V$ as nonmanipulators.

As benchmarks, over time
eleven ``standard'' types of control questions have 
emerged~\cite{bar-tov-tri:j:control,hem-hem-rot:j:destructive-control,fal-hem-hem-rot:j:llull}, 
each with a
constructive version
and a destructive version.
Four of the eleven are each of
adding/deleting (at most $k$, with $k$ part of the input)
candidates/voters.  A fifth is so-called unlimited adding of candidates.
The remaining six are partition of candidates, runoff partition of
candidates, and partition of voters, each in both the model where
first-round ties promote and in the model where first-round ties
eliminate.
The
benchmark types are modeled on situations from the real world.  For
example, constructive control by adding voters models highly targeted
get-out-the-vote drives and constructive control by adding candidates
models trying to draw into elections spoiler candidates who will hurt
your opponents.  See for
example~\cite{bar-tov-tri:j:control,hem-hem-rot:j:destructive-control,fal-hem-hem-rot:j:llull} for 
a detailed discussion of the motivations of the benchmark control types.
We mention in passing that recent
  work~\cite{hem-hem-men:c:search-versus-decision} has shown that in
  the nonunique-winner model two pairs of the eleven destructive types are
  pairwise identical as to which instances they can be carried out on,
  while one pair is identical in the unique-winner model.

Due to there being many control types, giving their definitions can
take up a large amount of space.  In order to keep the paper
self-contained for those who want more formal definitions, as
Appendix~\ref{sec:control-definitions} we provide in their typically
stated forms the definitions of all the control types.  For those
who already know the definitions or at this point simply want a 
brief treatment (knowing Appendix~\ref{sec:control-definitions} 
is available with
formal definitions), the rest of the 
present paragraph
gives a short overview of the flavor of the different benchmark control 
attacks.
All these
problems have as their input an election, $(C,V)$, and a distinguished
candidate $p\in C$.  Constructive (destructive) control by deleting
voters---for a given election system, of course---also has a nonnegative
integer $k$ in the input and asks whether there is a subset of $V$ of
cardinality at most $k$ such that with that subset removed $p$ is (is not) a
winner.  Control by adding voters is analogous, except the input is
the election, $k$, and a set $W$ of voters who can be added (but at
most $k$ can be added).  
Deleting candidates and adding candidates are
analogous to the voter cases, with a $k$ as part of the
input, and the only twist is that in destructive control by
deleting candidates, it is forbidden to delete $p$.  
Unlimited adding of candidates is the same except
there is no limit $k$.  
Constructive
(destructive) partition of voters, in the ties-promote model, asks whether
there is a way of partitioning the voters into two groups so that if
all winners under the election system of each of those first-round
elections compete in a final election under the same election system
in which all voters vote (with their votes masked down to the
remaining candidates), $p$ is (is not) a winner.  In its
ties-eliminate variant, only unique-winners of a first round election
move forward.  The runoff partition of candidates types are
analogous, except in the first round it is the candidates that are
partitioned and all voters vote in each of those subelections.
Partition of candidates has just one side of the partition participating
in the first-round election, while the others get a bye to the final
round.

A problem is said to belong to the class FPT (is said to be 
fixed-parameter tractable, see~\cite{nie:b:invitation-fpt}) with respect to
a parameter if there is an algorithm for the problem whose
running time is $f(j)I^{O(1)}$, where $j$ is the input's value of that
parameter, $f$ is a computable function, and $I$ is the input's size.
Note that this means that although the algorithm for larger
values of $j$ can have a bigger multiplicative constant, the degree of
the polynomial running time is uniformly bounded from above---there is
some single integer $k$ such that regardless of the fixed $j$ the
algorithm for that parameter bound runs in time $O(I^k)$.  
Our parameter will almost always be the most natural one---the number
of candidates. In Section~\ref{s:weighted}, we will have cases where
our parameter is a tuple of features of the input rather than a single
feature, i.e., is a ``combined'' parameter
(see~\cite[Chapter~9]{bet:thesis:multivariate}).  
We stress that FPT should not be confused with the class, of which it
is a subset, XP\@.  XP problems are in P for each fixed bound on the
parameter, but their degree can grow as that bound grows.  Thus XP is
a far less attractive class, even though by brute force it is often
clear that parameterized versions of voting problems fall into it.  
However, our focus is firmly on  the more demanding
goal of establishing FPT results.

\section{Related Work}\label{s:related-work}
The computational complexity of manipulation, bribery, and control for
Schulze voting and ranked pairs has been studied previously by Parkes
and Xia, Xia et al., Menton and
Singh, and Gaspers et 
al.~\cite{par-xia:c:ranked-pairs,con-pro-ros-xia-zuc:c:unweighted-manipulation,men-sin:c:schulze,gas-kal-nar-wal:c:schulze}.
The work of Xia et al.\ and Parkes and Xia establishes
(see also the table
in~\cite{par-xia:c:ranked-pairs})
that for Schulze elections
constructive and destructive bribery, constructive and destructive
control by adding and deleting voters, and constructive control by
adding candidates are NP-complete, and that ranked pairs has not
only all these hardnesses but also has NP-completeness results for
constructive and destructive manipulation, for destructive control by
adding candidates, and for constructive and destructive control by
deleting candidates.  
Menton and Singh (see Table 1
of~\cite{men-sin:c:schulze}) studied all remaining types of
constructive and most types of destructive control for Schulze voting,
and their work 
establishes that NP-completeness holds additionally for constructive
control by unlimited adding of candidates, constructive control by
deleting candidates, each variant of constructive control by partition
or runoff partition of candidates, and each variant of constructive
and destructive control by partition of voters,
and for the four cases of destructive control by partition 
or runoff partition of
candidates they show that polynomial-time algorithms
exist.\footnote{For Schulze, three of the 22
benchmark control cases---each 
clearly belonging to $\np$ for Schulze---were left open by 
the abovementioned papers and 
indeed even now these cases remain open as to whether they
are in P, are NP-complete, or have some other complexity:
destructive control by 
adding candidates, destructive control by 
deleting candidates, 
and 
destructive control by 
unlimited adding of candidates.  However, for each of those 
three (and all other of the benchmark control cases), we obtain
membership in FPT\@.}
The (unweighted coalitional) manipulation problem 
is shown to be in P for the 
destructive case by Parkes and 
Xia~\cite{par-xia:c:ranked-pairs} and for the constructive case by 
Gaspers et al.~\cite{gas-kal-nar-wal:c:schulze}.
Gaspers et al.~\cite{gas-kal-nar-wal:c:schulze}
prove that the \emph{weighted} constructive 
coalitional manipulation problem 
for Schulze elections is in polynomial time for 
each fixed number of candidates.  We observe that inspection
of their 
paper immediately makes clear that they indeed have 
even established the stronger claim that the 
weighted constructive coalitional manipulation problem for 
Schulze elections is in the class FPT\@.\footnote{Briefly, 
the reason their algorithm for 
the weighted constructive coalitional manipulation problem
  for Schulze elections is clearly even an FPT algorithm is as follows.
Their algorithm is using 
the fact, observed independently
by Menton and Singh~\cite{men-sin:t1ShowVersion:schulze}
and 
Gaspers et al.~\cite{gas-kal-nar-wal:c:schulze},
that for 
the nonunique-winner model, weighted 
constructive coalitional manipulation problem
  for Schulze elections, if one can make a given candidate a winner
  then there is a set of manipulative votes \emph{in which all
    manipulators vote the same way} and that candidate is selected 
  as a winner.  Once one has this, an FPT algorithm is obtained 
  simply by cycling over all possible preference orders, for each
  seeing whether, if that is what all the manipulators cast as their
  vote, the given candidate becomes a winner.  And that is precisely
  what their short, elegant algorithm is doing for this case.}

All the results in the two
papers involving Xia are in the unique-winner model.   Ranked pairs is
resolute (has exactly one winner), as Parkes and Xia frame it, and we follow
their framing.  And so the nonunique-winner and the unique-winner
models are in effect the same for ranked pairs.  Schulze is not resolute, but
although Parkes and Xia's
results on that are in the unique-winner model, they
comment that their results all also hold in the nonunique-winner
model.  Gaspers et al.\ study both the nonunique-winner model 
and
the unique-winner model.  
Menton and Singh use the nonunique-winner model as their basic
model, as do we in the present paper.  To us the nonunique-winner model
is more attractive in
not requiring a tie-breaking that, especially in symmetric cases, is
often arbitrary and can change the flavor of the system.  However, our
main FPT results are proven by a loop approach over 
ILPFPs (integer linear programming feasibility problems), and it
is clear that a straightforward adjustment to these will also handle
the unique-winner cases.  The key difference between our work and all the
abovementioned work is that our work is in general looking at the
complexity of these problems when parameterized by the number of
candidates, and for this we give FPT algorithms.  The earlier
papers primarily looked at unbounded numbers of candidates and obtained both 
P and NP-completeness results;  our contribution is that for all their
NP-complete cases, we show membership in FPT\@.

As to technique, the closest precursors of this paper are two papers
by Faliszewski et
al.~\cite{fal-hem-hem-rot:j:llull,fal-hem-hem:j:multiprong}.  Those,
like us, use a loop over ILPFPs.  The main differences between that
work and ours is that (a)~they deal with control, and we also are
concerned with bribery and manipulation, and (b)~as explained in
detail in Section~\ref{ss:key}, their type of loop-over structure
isn't flexible enough for our cases, and the natural structure for us
to loop over generates a number of objects not bounded in the number
of candidates, and so in this paper we find a middle ground that
allows the loop to be over a bounded-in-$\|C\|$ number of objects yet
provides enough information in the objects so as to allow the ILPFPs
to complete the checking of whether success is possible.  For a
different type of attack 
known as ``swap bribery'' and a different 
election system, 
Dorn and
Schlotter~\cite{dor-sch:j:parameterized-swap-bribery} 
have recently employed what in
effect (although implicitly) is a loop over ILPFPs, 
and they mention in passing without 
details that that swap bribery 
approach should apply to ranked pairs.\footnote{Dorn and
Schlotter~\cite{dor-sch:j:parameterized-swap-bribery} 
separately
mention in passing and without definitions or 
details that their 
approach should apply to manipulation and some unspecified variants of
control.  If one takes as implied there the combination of their
ranked-pairs aside and their other attacks-aside (and they in their
paper don't explicitly assert that),
then that 
without-details assertion pair, combined, overlaps some of our
ranked-pairs results regarding control (though they do not specify
which control types they are speaking of).
However, in contrast, we here actually provide a certification
framework handling the ranked pairs election system.  We 
believe that 
their results, even if looking at the combination of their 
asides, don't overlap our main work on manipulation, since
they speak of ``weighted and unweighted manipulation,'' 
but by that undefined-there use they seem to mean
a weaker notion of manipulation than that used in the 
present paper, namely, we are looking at coalitional 
manipulation, but they seem to be referring to 
noncoalitional manipulation.  (The reason we say this is that
for the weighted case of coalitional manipulation, the natural
ILPFPs one would generate have neither their number of 
variables nor their number of constraints ``bounded in
the number of candidates independently of the number of voters.''
Beyond this, for ranked pairs we will prove, as Theorem~\ref{ranked-pairs-hard},
 that weighted coalitional manipulation is NP-complete for each fixed number of candidates 
starting at five; 
 this easily implies that unless $\pisnp$ no FPT algorithm can exist.
There we will discuss how far toward such algorithms
one can seem to 
get within the ILPFP approach and its extensions, namely, by looking 
at the special case of bounded weights and even of 
bounded weight-set-cardinality. That section notes
that we can handle even the coalitional case for bounded weights,
and indeed even for unbounded weights but 
bounded manipulator-weight-set-cardinality, and that 
latter case itself is a generalization of 
weighted noncoalitional manipulation.)}

Taking an even broader perspective, this work is part
of a line that looks at the complexity of elections in the context of
bounds on the number of candidates, a study that for example has been
pursued famously by Conitzer, Sandholm, and
Lang~\cite{con-lan-san:j:when-hard-to-manipulate} regarding at what
candidate numbers complexity jumps from P to NP-complete.  The particular
focus on FPT algorithms, and maintaining a uniform degree bound over
all values of bounds on the number of candidates, is part of the
important field of parameterized complexity (see~\cite{nie:b:invitation-fpt}, 
see~\cite{bet-bre-che-nie:c:parameterized-elections-survey} for a survey on 
this approach for elections, and 
see~\cite{rot-sch:j:typical-case-challenges} for a survey of 
an alternate 
approach to bypassing complexity results).
\section{Results by Looping over Frameworks}\label{s:loops}

We now present our results that are established 
by our looping-over-frameworks idea.
We will handle in separate sections bribery, control, and manipulation,
showing how to achieve FPT results for each.  

In the bribery section, 
\sectionref{ss:bribery},
we will first prove the bribery result for Schulze elections, 
so that the reader quickly gets to seeing how the proof goes without having to have first 
seen how the approach works for ranked pairs.
We then will give our ranked pairs winner-set certification framework, 
and will note how to convert our proof into a proof for that case also. 
Then later in the control section,
\sectionref{ss:control}, we will state and prove together the 
Schulze-elections 
case and the ranked-pairs case.

Since
(unweighted) Schulze manipulation has been 
shown 
to be in P in general, both 
for the constructive~\cite{gas-kal-nar-wal:c:schulze} 
and the destructive~\cite{par-xia:c:ranked-pairs} 
cases, we do not need to handle 
Schulze elections in our manipulation section, \sectionref{ss:loop-manip}.
(For the
nonunique-winner model, 
weighted constructive coalitional manipulation for 
Schulze elections, 
Gaspers et al.~\cite{gas-kal-nar-wal:c:schulze} 
provide what as mentioned earlier is an 
FPT algorithm.  And in
Section~\ref{s:weighted} 
we will show, as part of Theorem~\ref{t:w-manip-card},
that for Schulze elections 
our approach provides an FPT algorithm for
special 
cases of 
both
weighted 
\emph{destructive} coalitional manipulation
and 
unique-winner model, 
weighted constructive coalitional manipulation.)

\subsection{Bribery Results and Specification of Ranked Pairs Winner-Set Certification Framework}\label{ss:bribery}

In this section, we will first state and prove the bribery 
result for Schulze, then will give our winner-set certification
framework for ranked pairs, and then will state and prove our 
bribery result for ranked pairs.

\subsubsection{Bribery Result for Schulze}\label{sss:bribery-schulze}
We state and prove the bribery result for Schulze.

\begin{theorem}\label{t:bribery-schulze}
  For Schulze elections, bribery is in FPT 
  (is fixed-parameter tractable) with respect to the
  number of candidates,
  in both the succinct and nonsuccinct
  input models, for both constructive and destructive bribery, in both
  the nonunique-winner model and the unique-winner model.
\end{theorem}

\begin{proofs}
We first give the proof for the constructive, nonunique-winner model 
case.  We will handle simultaneously the succinct and nonsuccinct cases.

Our FPT algorithm works as follows.  It gets as its input an instance
of the bribery problem, and so gets the candidates (with a
distinguished candidate noted), the votes (or for
the succinct version, a list of which types of votes occur at least
once, along with the multiplicities of each), 
and the limit $k$ on how many voters can be bribed.
Let $j$ be the number of candidates in the input instance.
To mesh with the naming scheme within our SWCFs, we immediately rename
all the candidates (including within the votes) to be
$1,\ldots,j$, with the distinguished
candidate becoming candidate 1.  Now, the top-level programming 
loop of the algorithm is as specified in 
Algorithm~\ref{alg:loop-bribery} 
(recall that we are showing the constructive,
nonunique-winner model case in this algorithm).
\begin{algorithm} 
\caption{Top-level loop for bribery}
\label{alg:loop-bribery}
\begin{algorithmic}
\STATE \noindent{\bf Start}
\FOR{each $j$-SWCF $K$}
\IF{candidate 1 is a winner according to $K$ and $K$ is an
internally consistent, well-formed $j$-SWCF}
\STATE (1) build an ILPFP that checks whether there is a way of 
bribing at most $k$ of the voters such that $K$'s winner-set 
certification framework is realized by that bribe
\STATE (2)  run that ILPFP and if it can be satisfied then halt and 
accept (note: the satisfying settings will even let us output the precise
bribe that succeeds)
\ENDIF
\ENDFOR
\STATE declare that the given goal cannot be reached by using at most 
$k$ bribes
\STATE {\bf End}
\end{algorithmic}
\end{algorithm}

\smallskip
\noindent

All that remains is to specify the ILPFPs that we build 
inside the loop, for each given $j$-SWCF $K$.  Suppose 
we are doing that for some particular $K$.  We do it as follows.

There are $j!$ possible votes over $j$ candidates; let us 
number them from $1$ through $j!$
in any natural computationally simple-to-handle way.  We call the $i$th of these 
the $i$th ``vote type.''
We will have constants $n_i$, $1 \leq i \leq j!$, denoting how 
many voters start with vote type $n_i$.  Our ILPFP will have 
integer variables (which we will ensure are nonnegative) 
$m_{i,\ell}$, $1 \leq i \leq j!$, $1 \leq \ell \leq j!$.
$m_{i,\ell}$ is the number of voters who start with vote type $i$ but 
are bribed to instead cast vote type $\ell$.  (Having 
$m_{i,i} \neq 0$ is pointless but 
allowed, as is having 
simultaneously $m_{i,\ell}>0$ and $m_{\ell,i}>0$.)
So the number of ILPFP variables is $(j!)^2$, which is large 
but is bounded with respect to $j$, so 
Lenstra's algorithm~\cite{len:j:integer-fixed} 
can be used to deliver an FPT performance overall.

Also, not as direct parts of the ILPFP but as tools to help us build
it, we define two boolean predicates, $\mathit{Bigger}(a,b,c,d)$ and
$\mathit{StrictlyBigger}(a,b,c,d)$, where the arguments each vary over
$1,\ldots,j$.  
Let us use $\netadv(a,b)$ to indicate the weight of the 
WMG edge (after our manipulative actions) that points from $a$ to $b$.
Recall, from Section~\ref{ss:key}, that what a 
\mbox{$j$-SWCF} does---and so in particular 
what our 
under-consideration 
\mbox{$j$-SWCF}, $K$, does---is 
specify (for a very large number of such quadruples, though that
number actually is bounded as a function of $j$ though we do not
need that since our number of variables is already bounded as a function
of $j$) that $\netadv(a,b) \geq \netadv(c,d)$ or that $\netadv(a,b) > \netadv(c,d)$, i.e.,
that in the WMG, a certain edge is greater than or equal to another
edge in weight, or is strictly greater in weight.  
For each such specified relation that
explicitly appears in $K$, set to true that bit in the appropriate
predicate ($\mathit{Bigger}$ or $\mathit{StrictlyBigger}$), 
and leave all the other bits set to false.  We of course 
will have to enforce these specifications through our constraints.

\begin{figure}[!tp]
$\Bigg(
\displaystyle\sum_{i\in \pref(a,b)}
\bigg(n_i -  \big(\sum_{1\leq \ell' \leq j!} m_{i,\ell'}\big) + 
\big(\sum_{1\leq i '\leq j!} m_{i',i}\big) 
\bigg)
\Bigg)$~~~~~~~~~~~~~~~~~~~~~~~~~~~~~~~~~~~~~~~~~~~~~~~~~~~~~~~~~~~~\\
{\centering
$${}~-~{} 
\Bigg(
\sum_{i\in\pref(b,a)}
\bigg(n_i -  \big(\sum_{1\leq \ell' \leq j!} m_{i,\ell'}\big) + 
\big(\sum_{1\leq i' \leq j!} m_{i',i}\big) 
\bigg)
\Bigg)
$$}\\%
\mbox{~$\geq 1{}~+~{}
\Bigg(
\displaystyle\sum_{i\in\pref(c,d)}
\bigg(n_i -  \big(\sum_{1\leq \ell' \leq j!} m_{i,\ell'}\big) + 
\big(\sum_{1\leq i' \leq j!} m_{i',i}\big) 
\bigg)
\Bigg)$}~~~~~~~~~~~~~~~~~~~~~~~~~~~~~~~~~~~~~~~~~~~~~~~~~~~~~~~~~~~~~~~~~\\
{\centering
$${}~-~{} 
\Bigg(
\sum_{i\in\pref(d,c)}
\bigg(n_i -  \big(\sum_{1\leq \ell' \leq j!} m_{i,\ell'}\big) + 
\big(\sum_{1\leq i' \leq j!} m_{i',i}\big)
\bigg)
\Bigg).$$}
\caption{\label{f:big-equation}Constraint enforcing that, after 
the bribes happen, $\netadv(a,b) > \netadv(c,d)$.}
\end{figure}

Now we can specify all the constraints of our ILPFP\@.  There will
be three types of constraints.  The first are the housekeeping 
constraints to make sure that the number of bribes and 
the $m_{i,\ell}$s are all reasonable. Our constraints of this sort are:
For each $1\leq i,\ell\leq j!$, we have a constraint $m_{i,\ell} \geq 0$.
For each $1\leq i \leq j!$ we have a constraint 
$n_i \geq \sum_{1\leq z \leq j!} m_{i,z}$;
that is, we do not try to bribe away from vote type~$i$ more votes than
initially exist of vote type~$i$.
And we have the constraint
$k \geq \sum_{1\leq i',\ell'\leq j!} m_{i',\ell'}$; 
that is, our total number of 
bribes does not exceed the bribe limit~$k$. 

The second type of constraint consists of those 
constraints used to enforce the bits set 
to ``true'' in $\strictlybigger$.  For each such bit, we will
generate one constraint:  for a bit that is 
saying that $\netadv(a,b) > \netadv(c,d)$, we will enforce 
that with the constraint shown in 
Figure~\ref{f:big-equation};
in that figure and for the rest of this paper,
in order to make the representation 
of the constraints more concise,
we introduce the shorthand
notation $\pref(a,b)$ as the set of vote types $i$, $1 \leq i \leq
j!$, in which candidate $a$ is preferred to candidate $b$.
All that the bulky-looking constraint of the figure says is that 
after all the gains and losses due to bribing happen, 
the number 
of voters who prefer $a$ to $b$ minus the number who prefer 
$b$ to $a$ is strictly larger than 
the number 
of voters who prefer $c$ to $d$ minus the number who prefer 
$d$ to $c$.  
If $\strictlybigger(a,b,c,d)$ is set to ``false,''
that does \emph{not} mean we generate a constraint ensuring 
that $\netadv(a,b)\not>\netadv(c,d)$.  
Rather, if a given 
bit is set to ``false,'' that just means that that particular bit-setting
does not itself create a constraint.
In contrast,
bits set to ``true''
in $\strictlybigger$ and $\bigger$ mean that we generate a 
constraint to enforce the stated relation.

The third type of constraint consists of those 
constraints used to enforce the bits set 
to ``true'' in $\bigger$.  For each such bit, we will
generate one constraint: for a bit that is 
saying that $\netadv(a,b) \geq \netadv(c,d)$, we will enforce 
that with precisely the constraint shown in 
Figure~\ref{f:big-equation}, except with the ``1+'' removed from
the right-hand side of the inequality.

That completes our statement of the ILPFP, which indeed captures
what it seeks to capture.  And using 
Lenstra's algorithm~\cite{len:j:integer-fixed}
for each of
our ILPFPs, the overall loop over the ILPFPs has the desired running
time.  (Although for each fixed $j$ the multiplicative constant is
very large, the degree of the polynomial, which is uniform over all $j$, 
isn't terrible; Lenstra's algorithm uses just a linear number of 
arithmetic operations on linear-sized integers~\cite{nie:b:invitation-fpt}.
Still, even within the good news that we have placed the problem within FPT,
there is the bad news that the multiplicative constant is so large that 
this FPT algorithm does not provide an algorithm for practical
use.)  To be clear, since what is a constant and what is a 
variable is a bit subtle here, let us say a bit more about the use of 
Lenstra here.  
What we in effect are using is that 
Lenstra's work ensures that there is a $k$ such that
for each fixed number of candidates and 
each of the (large but bounded as a function
of the number of candidates) ILPFPs generated in our loop, if we view
that ILPFP as an object whose running time for solution is
being evaluated asymptotically as the number of voters
increases without bound, the ILPFP's running time is $\bigo(n^k)$.  
(That same value $k$ holds 
for all numbers of candidates and for 
all ILPFPs that our loop generates for that number of candidates.  
However, for different numbers of candidates the multiplicative constant 
represented by the ``big O'' may differ.)
Note that each such ILPFP
object in effect has as \emph{its} set of variables (regarding 
the asymptotics of its running time) the 
\emph{constants} of the ILPFP;  and a big part of 
what our looping algorithm does 
is to set those constants based on the votes in the election.

That 
was the proof for the 
constructive, nonunique-winner model 
case.
To change the above proof from the constructive to the destructive
case and/or from the nonunique-winner case to the unique-winner
case, in the main loop we will simply create ILPFPs for only those SWCFs
whose set of who wins and loses reflects a sought outcome.  
For example, for
the destructive case in the unique-winner model, that would be having
the distinguished candidate not be a unique winner, i.e., 
the start of 
Algorithm~\ref{alg:loop-bribery}'s ``if'' statement would become
``if candidate 1 is not a unique winner.''~\end{proofs}

The same proof approach 
applies to ranked pairs. However, we first must do 
some 
work to define an appropriate
winner-set certification framework for ranked pairs.
We turn to that now.
\subsubsection{Specification of Ranked Pairs Winner-Set Certification Framework}
\label{s:ranked-pairs}
In this section, we describe the 
winner-set certification
framework that makes our approach work for ranked pairs.

Basically, an instance of that framework will be a story that tells us
what happens at each stage of the iterative process that defines
ranked pairs.  We could actually tell this story without fixing up
front, for 
each pair $\{a,b\}$  
of distinct candidates,
whether $a$ 
is preferred to $b$ by a
majority of the voters, 
or whether 
$b$ 
is preferred to $a$ by a
majority of the voters, 
or whether $a$ and $b$ exactly tie as to how many voters 
prefer one to the other.  Not fixing that information up front 
would improve our multiplicative factor that depends on but is 
fixed for each fixed number of candidates.  But we are not focused on
that factor.  So to make things particularly simple to describe, we
are here just going to toss into our framework a fixing of all such
pairwise-outcomes-in-the-WMG\@.  As in the Schulze case, we will have
changed all the names of the candidates to be 1 through $\|C\|$ (and
will have remapped our tie-breaking function in the same way).  So,
one part of our framework is, for each (unordered) pair of distinct
candidates
$\{a,b\}$, a claim as to which one of these holds: the WMG edge from $a$ to
$b$ is strictly positive, the edge from $b$ to $a$ is strictly
positive, or both edges are $0$.  (We do not include any claim about
the precise value of those edge weights; that would create a framework
whose number of instances, for a fixed number of candidates, grew with
the number of voters---something 
we must firmly avoid.)  And an instance of the
framework then goes step by step through the process the ranked-pairs
algorithm goes through, but in a somewhat ghostly way in terms of what
it specifies.  For each step of the process, it makes a claim as to
what pair of candidates is considered next, and a claim as to whether
that pair of candidates will be skipped permanently due to it having
been already set (due to transitivity) by earlier actions of our flow
through ranked pairs (that isn't an on-the-fly thing in that the
instance itself has all its earlier claims and so we can even make
sure to loop only over instances of the framework that are internally
consistent regarding this), and if it is not skipped, a claim about
which of the two outcomes happens (which is placed above the other in
our ranked-pairs outcome; again, we can read this from those
choice-of-3-possibilities settings we did up front, plus the feasible
tie-breaking if needed).  So that is the story the framework provides,
and a given instance of the framework will (if properly formed) set an
ordering over all the candidates.
As before, the 
algorithms will loop over instances of these frameworks, doing so 
over only instances that have the desired outcome (e.g., ``$p$ is a unique
winner'') and that aren't obviously internally inconsistent. 
(For our
partition by voter cases, there is a double-loop over such frameworks,
to handle both subelections.)

As in the Schulze case, we
will use the ILPFPs to see if the given kind of control can create a
case where the given framework can be made to hold.  All the
housekeeping work in the ILPFPs as to tallying how the votes are
bribed/controlled/manipulated is still needed here (so the variable
sets are the same as the ones for Schulze).  But note, crucially, that we now must
enforce not things about paths, but rather we must enforce that 
the framework's guesses about whether the edge from $a$ to $b$
is negative, positive, or 0 after the bribery/control/manipulation are
all correct (this is very natural to enforce with constraints, within
the ILPFP framing), \emph{and must also enforce that the framework's claim
about which candidate pair is considered next is what would actually
happen under the votes that emerged from the
bribery/control/manipulation}.  But that latter claim, for each step in
the story, can be checked by appropriate, carefully built
constraints, written with close attention paid to the tie-breaking rule
among pairs.  These constraints will be pretty much our favorite sort
of constraint---seeing whether a WMG edge is greater than or equal to
another, or seeing whether it is strictly greater than another.  
This will be made clearest
by an example. 
Suppose our candidates are named \textit{1}, \textit{2}, 
\textit{3}, and \textit{4}.  And suppose the 
tie-breaking order on unordered 
pairs is 
$\{\textit{4},\textit{3}\}>\{\textit{4},\textit{2}\}>
\{\textit{4},\textit{1}\}>\{\textit{3},\textit{2}\}>
\{\textit{3},\textit{1}\}>\{\textit{2},\textit{1}\}$,
and on candidates is $\textit{4}>\textit{3}>\textit{2}>\textit{1}$.
(This is exactly a case of the 
sample feasible rule pair we gave in 
Section~\ref{s:defs}, where for example tied pairs are tie-broken 
  in favor 
  of the pair with the lexicographically-larger larger-candidate-of-the-pair,
  and when the larger members
  are the same in both pairs then the tie is broken in favor of 
  whichever pair has the lexicographically-larger 
  smaller-candidate-of-the-pair.  
 We've written our unordered pairs, in the tie-breaking order above,
with the lexicographically larger element first simply to make it clear
why Section~\ref{s:defs}'s 
example tie-breaking rule would put them in the order shown above.)
Suppose the RPWCF 
says the first pair to be compared is $\{\textit{1},\textit{4}\}$
and that the outcome is $\textit{4} \succ \textit{1}$.  
Let $\netadv(a,b)$ be defined as before.
To check that 
$\textit{4} \succ \textit{1}$ is the right outcome, since 
$\textit{4}>\textit{1}$ in the tie-breaking
function we need to check that $\netadv(\textit{4},\textit{1}) 
\geq \netadv(\textit{1},\textit{4})$
(if $\textit{1} \succ \textit{4}$ in the tie-breaking order, we'd check that 
$\netadv(\textit{4},\textit{1}) \geq 1+\netadv(\textit{1},\textit{4})$);  
we can read this right off
the 
3-way-claim as to how \textit{1} and \textit{4} compare in their
head-to-head contest, which itself we'll enforce in constraints.  
And as to the claim that $(\textit{1},\textit{4})$ was the first pair to be 
compared, in light of the tie-breaking order, that can be enforced
using 10 constraints: 
6 saying that our pair ties or beats those below us 
in the tie-breaking 
ordering ($\netadv(\textit{4},\textit{1}) \geq \netadv(a,b)$
for the $(a,b)$ values 
$(\textit{3},\textit{2})$,
$(\textit{2},\textit{3})$,
$(\textit{3},\textit{1})$,
$(\textit{1},\textit{3})$,
$(\textit{2},\textit{1})$,
$(\textit{1},\textit{2})$), and 
4 saying that our pair strictly beats those above us 
in the tie-breaking ordering ($\netadv(\textit{4},\textit{1}) \geq \netadv(a,b)$
for the $(a,b)$ values 
$(\textit{4},\textit{3})$,
$(\textit{3},\textit{4})$,
$(\textit{4},\textit{2})$,
$(\textit{2},\textit{4})$).
We could cut those $6+4$ constraints to $3+2$ if we wish, by using the 
value of the 3-way-claim for  each of those 5 other pairs.
Note that all these comparisons are about post-bribe/manipulation/control
vote numbers---things we do know how to easily put into an ILPFP 
constraint, basically, by appropriate summations.
Moving on, if the framework says the next unordered pair 
after $\{\textit{1},\textit{4}\}$
to be considered is $\{\textit{1},\textit{3}\}$ and that 
the outcome is $\textit{1} \succ \textit{3}$, we of course will not need to 
enforce any comparisons with 
$\netadv(\textit{4},\textit{1})$ 
or $\netadv(\textit{1},\textit{4})$;  we will
generate and put into the ILPFP just the needed/appropriate comparisons.
If the framework after that says the third pair to consider 
is $\{\textit{3},\textit{4}\}$ but it also says that 
(due to $\textit{4}\succ \textit{1} \succ \textit{3}$ already being set)
the pair $\{\textit{3},\textit{4}\}$ gets skipped, we under 
our RPWCF framework still must generate the constraints
to check that $\{\textit{3},\textit{4}\}$ truly under our votes 
as they now are
\emph{did} deserve to come up next (we mention in passing that 
we could skip that check as long as we adjust our 
ILPFP to not check anything regarding pairs that 
are already related, even transitively, under the $\succ$'s so far---a
slightly different approach than ours but also quite fine), but the generated
comparisons-to-check-that will not do comparisons against things the
existing order so far ($\textit{4}\succ \textit{1} \succ \textit{3}$) 
takes out of play.  This
completes our description of our RPWCF notion.

\subsubsection{Bribery Result for Ranked Pairs}
Having specified the ranked pairs winner-set certification framework,
the bribery case for ranked pairs 
can now be stated and justified.

\begin{theorem}\label{t:bribery-ranked-pairs}
  For ranked pairs (with any feasible tie-breaking functions), bribery is in FPT 
  (is fixed-parameter tractable) with respect to the
  number of candidates,
  in both the succinct and nonsuccinct
  input models, for both constructive and destructive bribery, in both
  the nonunique-winner model and the unique-winner model.
\end{theorem}

\begin{proofs}
We use the programming loop of Algorithm~\ref{alg:loop-bribery} to 
now loop over not $j$-SWCFs, but instead over 
$j$-RPWCFs, with $j$
again being the number of candidates.
The variables of the ILPFPs will be the same as those for Schulze.
As for the constraints of the ILPFPs, all the housekeeping constraints for bribery remain intact.
Additionally an RPWCF, as described in Section~\ref{s:ranked-pairs}, guesses for each edge whether it is positive, negative, or zero in weight. We can easily handle these possibilities 
with the constraints
$\netadv(a,b) \geq 1$, $\netadv(b,a) \geq 1$, and $\netadv(a,b)=0$ respectively.
The other constraints enforced by an RPWCF are those between pairs of edges,
and they can be successfully captured by the $\strictlybigger$ and $\bigger$ predicates 
defined in the proof of Theorem~\ref{t:bribery-schulze}. 
Thus we clearly can build an ILPFP encoding an RPWCF and the constraints of the
bribery problem in the same way we did for Schulze elections, and
bribery is in FPT for ranked pairs as well.~\end{proofs}

\subsection{Control Results}\label{ss:control}

In this section, we prove our results for control gained through the
looping-over-frameworks approach.
Our proofs for these cases will closely follow our general
looping-over-frameworks structure, and we will just have to appropriately
build the constraints of our ILPFPs to  
handle the details of the control problems.

\begin{theorem}
\label{t:adv-param-c-a}
  For Schulze elections and for (with any feasible tie-breaking
  functions) ranked pairs, 
 control by adding voters
 is in FPT with respect to the
  number of candidates,
  in both the succinct and nonsuccinct input models,
  for both constructive and destructive control,
  and in both the nonunique-winner model and the
  unique-winner model.
\end{theorem}

\begin{proofs}
We will handle together all the cases of this theorem.

  Again, let the candidates in the control problem be $1,\ldots,j$
  with the distinguished candidate being 1. The control problem has as
  its input a set of initial votes $V$ and a set of additional
  votes $W$ (or for the succinct version, one list for each of these two sets describing
  which types of votes occur at least once in that set, along with the
  multiplicities of each), the latter of which contains the
  votes that can be added by the control action. Also there is a limit $k$
  on the number of votes that can be added from $W$.

The top-level programming loop is as described in Algorithm~\ref{alg:loops-control-add-v}
(with WCF being either SWCF for Schulze 
or RPWCF for ranked pairs).

\begin{algorithm}
\caption{Top-level loop for control by adding voters}
\label{alg:loops-control-add-v}
\begin{algorithmic}
\STATE \noindent{\bf Start}
\FOR{each $j$-WCF $K$}
\IF{candidate 1 
(is/is not) (a winner/a unique winner) 
according to $K$ and $K$ is an
internally consistent, well-formed $j$-WCF}
\STATE (1) build an ILPFP that checks whether there is a set of
votes $W'\subseteq W$, with $\|W'\| \leq k$,
 such that $K$'s winner-set 
certification framework is realized by the set of votes $V \cup W'$
\STATE (2)  run that ILPFP and if it can be satisfied then halt and 
accept (note: the satisfying settings will even let us output the precise
added set that succeeds)
\ENDIF
\ENDFOR
\STATE declare that the given goal cannot be reached by adding at most $k$ voters
\STATE {\bf End}
\end{algorithmic}
\end{algorithm}
\smallskip
\noindent

The two binary selections in the algorithm's ``if'' statement 
are made as in the proof of Theorem~\ref{t:manip-param-c}.

All we have to do now is show how we build the ILPFP inside the loop.
Again, as with manipulation, we will have two groups of constraints:
those corresponding to the WCF-enforcing predicates and those
corresponding to the structure of the control problem. For every $i$,
$1 \leq i \leq j!$, we will have a variable $v_i$ representing the
number of votes of type $i$ in $W'$ (i.e., 
the $W'$ sought by 
Algorithm~\ref{alg:loops-control-add-v}), a constant $n_i$ representing
the number of votes of type $i$ in $V$, and a constant $h_i$
representing the number of votes of type $i$ in $W$.

As we described in the proof of Theorem~\ref{t:manip-param-c},
if we can represent $\netadv(a,b)$ as a linear expression
in terms of our constants and variables, we can
implement all the WCF-enforcing constraints (those of the $\strictlybigger$ and 
$\bigger$ predicates and those of the 3-way possibilities for ranked pairs)
as linear constraints in our ILPFP\@. 
Thus, using the shorthand notation $\pref(a,b)$ as
described in Section~\ref{ss:loop-manip}, 
we express $\netadv(a,b)$ as follows:
\[ \sum_{i \in \pref(a,b)} {(n_i+v_i)} ~~- \sum_{i' \in \pref(b,a) } {(n_{i'}+v_{i'})}.\]

As for the constraints ensuring the validity of the control action, we first need to
ensure that for every type of vote, the number of votes of that type in $W'$
is bounded by the number in $W$. 
For every vote type 
$i$, $1 \leq i \leq j!$, the constraint $v_i \leq h_i$ will enforce this in our ILPFP\@.
We also make the following constraint 
to enforce the adding bound:
\begin{equation}
\label{eq:add-bound}
\sum_{1 \leq i \leq j!} {v_i} \leq k.
\end{equation}
Here again all of our variables have to be nonnegative, and thus we have 
the constraint $v_i \geq 0$ for every
$i$, $1 \leq i \leq j!$.

This suffices to describe how we can build a WCF-enforcing ILPFP for
the control by adding voters problem in a way appropriate for our
looping-over-frameworks technique.  Thus we have an algorithm that
will run in uniform polynomial time for every fixed parameter value, putting
this problem in FPT\@.~\end{proofs}

\begin{theorem}
\label{t:adv-param-c-d}
  For Schulze elections and for (with any feasible tie-breaking
  functions) ranked pairs, 
 control by deleting voters
 is in FPT with respect to the
  number of candidates,
  in both the succinct and nonsuccinct input models,
  for both constructive and destructive control,
  and in both the nonunique-winner model and the
  unique-winner model.
\end{theorem}
\begin{proofs}
We will handle together all the cases of this theorem.

The input for this problem is  a control instance with a set of votes $V$ 
over $j$ candidates $1, \ldots, j$, 
with candidate $1$ being the distinguished candidate, 
and with a bound $k$ on the number of votes to be deleted. 
 Algorithm~\ref{alg:loop-del-vot} specifies the top-level loop.

\begin{algorithm}
\caption{Top-level loop for control by deleting voters}
\label{alg:loop-del-vot}
\begin{algorithmic}

\STATE \noindent{\bf Start}
\FOR{each $j$-WCF $K$}
\IF{candidate 1 
(is/is not) (a winner/a unique winner) 
according to $K$ and $K$ is an
internally consistent, well-formed $j$-WCF}
\STATE (1) build an ILPFP that checks whether there is a subset of the
voters $V'$, with $\|V'\| \leq k$, 
such that $K$'s winner-set 
certification framework is realized by the set of votes $V - V'$
\STATE (2)  run that ILPFP and if it can be satisfied then halt and 
accept (note: the satisfying settings will even let us output the precise
deleted set that succeeds)
\ENDIF
\ENDFOR
\STATE declare that the given goal cannot be reached by deleting at most $k$ voters
\STATE {\bf End}
\end{algorithmic}
\end{algorithm}

The two binary selections in the algorithm's ``if'' statement 
are made as in the proof of Theorem~\ref{t:manip-param-c}.

For each $i$, $1 \leq i \leq j!$,
the ILPFP we build inside the loop will include a constant $n_i$, 
representing how many votes of vote type $i$ are in the initial
election.  We will include in the ILPFP variables $v_i$, $1 \leq i \leq j!$,
 representing the number of votes of type $i$
that are deleted.  With this new variable, we will 
formulate $\netadv(a,b)$ in the ILPFP as follows:

\[
\sum_{i \in \pref(a,b)} (n_i - v_i)  ~~ - 
\sum_{i' \in \pref(b,a)} (n_{i'} - v_{i'}).
\]

With this we can easily build the $\bigger$ and $\strictlybigger$
predicates we use to enforce the WCF, as well as the additional
constraints necessary for an RPWCF\@.

Additionally, we need to include constraints that ensure that the
control action chosen through the assignment to the variables is a
legal one.  We include a constraint $v_i \geq 0$ for each $i$, 
$1 \leq i \leq j!$, enforcing that we cannot delete a negative number of
voters of any type.  We include constraints $n_i \geq v_i$ for each
$i$, $1 \leq i \leq j!$, enforcing that we cannot delete more voters
of any type than there were in the first place.  And finally we add a
constraint $k \geq \sum_{1 \leq i \leq j!} v_i$, bounding the total
number of deletions by the bound $k$.  

The above modifications to our basic  approach embed
the problem of deciding what subset of voters to delete to satisfy a
given WCF into an ILPFP, and there will be only a constant number of
ILPFPs to test for each number of candidates.  Thus the algorithm
specified by the outer loop above is uniformly polynomial for every
fixed parameter value, putting this problem in FPT\@.~\end{proofs}

\begin{theorem}\label{t:pv-param-c}
  For Schulze elections and for (with any feasible tie-breaking
  functions) ranked pairs, control by
 partition of voters 
  is in FPT with respect to the
  number of candidates,
  in both the ties-eliminate and ties-promote models,
  in both the succinct and nonsuccinct input models,
  for both constructive and destructive control,
  and in both the nonunique-winner model and the unique-winner model.
\end{theorem}

\begin{proofs}
We will handle together all the cases of this theorem.

  The cases of this theorem
  can again be handled with our looping-over-frameworks
  approach, and the constraints and modifications necessary to
  properly implement the WCF-enforcing ILPFP are similar to the
  previous cases, though we have to loop over not just single WCFs but
  rather pairs of them, one for each part of the partition.  As before
  we will assume the candidate names are $1,\dots,j$ with $1$ as the
  distinguished candidate.  We specify the outer loop for these cases
  in Algorithm~\ref{alg:loop-part-vot}.

\begin{algorithm}
\caption{Top-level loop for control by partition of voters}
\label{alg:loop-part-vot}
\begin{algorithmic}

\STATE \noindent{\bf Start}
\FOR{each $j$-WCF $K_1$}
\FOR{each $j$-WCF $K_2$}
\IF{ $K_1$ and $K_2$ are
internally consistent, well-formed $j$-WCFs}
\STATE (1) build an ILPFP that checks whether there is a partition of the
voters $V$ into $V_1, V_2$, such that $K_1$'s winner-set 
certification framework is realized by the set of votes $V_1$ and $K_2$'s winner-set 
certification framework is realized by the set of votes $V_2$
\STATE (2)  run that ILPFP and if it can be satisfied, and if
candidate 1 
(is/is not) (a winner/a unique winner) 
in the election
with voters $V$ and the surviving candidates from $K_1$ and $K_2$ (according to
the tie-handling rule in use, namely ties-eliminate or ties-promote), 
 halt and 
accept (note: the satisfying settings will even let us output the precise
partition that succeeds)
\ENDIF
\ENDFOR
\ENDFOR
\STATE declare that the given goal cannot be reached by any partition of voters 
\STATE {\bf End}
\end{algorithmic}
\end{algorithm}
 
The two binary selections in the algorithm's step~(2)
are made analogously to the decision in the ``if'' line within
the proof of Theorem~\ref{t:manip-param-c}.

Now we must specify the new details of the partition-handling ILPFPs\@.
For each $i$, $1 \leq i \leq j!$, 
we will include a constant $n_i$ representing how
many votes of vote type $i$ are in the initial election and 
a variable $v_i$
representing how many votes of type $i$ are put into $V_1$ (i.e.,
the $V_1$ sought by Algorithm~\ref{alg:loop-part-vot}).

Though we must use two WCFs, we can implement each as
before, and the constraints for each of a given pair of WCFs 
will appear
together as part of 
a 
single ILPFP\@.   
Let us use the predicates $\bigger_1$ and
$\strictlybigger_1$ ($\bigger_2$ and $\strictlybigger_2$) to 
handle the constraints of WCF $K_1$ ($K_2$) over
the weights of its WMG edges, which we will denote by $D_1$ ($D_2$).   

$\bigger_1$ and $\strictlybigger_1$ can be formulated appropriately in
terms of $\netadv_1$, while $\bigger_2$ and
$\strictlybigger_2$ can be formulated in terms of $\netadv_2$.  The
extra constraints for ranked pairs will be implemented in terms of
$\netadv_1$ and $\netadv_2$ as appropriate, as well.  
We will now show how to handle $\netadv_1$ and $\netadv_2$.
$\netadv_1(a,b)$ denotes how many voters in $V_1$ prefer
candidate $a$ to candidate $b$ and we can formulate it as follows:

\[
\sum_{i \in \pref(a,b)} v_i  ~~ - \sum_{i \in \pref(b,a)} v_i.
\]

$\netadv_2(a,b)$ denotes how many voters in $V_2$ prefer $a$ to $b$,
and we formulate it as follows:

\[
\sum_{i \in \pref(a,b)} (n_i - v_i)  ~~ - \sum_{i \in \pref(b,a)} (n_i - v_i).
\]

Finally, we must add constraints to ensure the chosen partition is a legal
one.  We include a constraint $v_i \geq 0$ for every $i$, $1 \leq i \leq j!$, 
enforcing that a nonnegative number of voters of each type are
chosen for the first partition.  We include a constraint $n_i \geq
v_i$ for every $i$, $1 \leq i \leq j!$, enforcing that we do not take
more voters than exist of each type for the first partition.

The above modifications to our basic  approach embed
the problem of deciding what partition of voters to use to satisfy a
pair of WCFs into an ILPFP, and there will be only a constant number of
ILPFPs to test for each number of candidates.  Thus the algorithm
specified by the outer loop above is uniformly polynomial for every
fixed parameter value, putting this problem in FPT\@.~\end{proofs}

\subsection{Manipulation Results}\label{ss:loop-manip}

Unweighted manipulation in Schulze 
elections has recently been shown to be in P by 
Gaspers et al.~\cite{gas-kal-nar-wal:c:schulze} for the 
constructive case.
So, since the destructive case was itself handled even 
earlier by Parkes and Xia~\cite{par-xia:c:ranked-pairs},
there is no need to cover (unweighted) manipulation for Schulze 
elections in this paper.

We now present our FPT result for (unweighted) manipulation 
in ranked pairs elections.
Although for this case we could 
use the  
looping-over-framework approach 
to obtain an FPT algorithm, we instead obtain an FPT 
algorithm here simply
as a consequence of Theorem~\ref{t:adv-param-c-a}.

Very loosely put, the idea behind this proof is that, in a certain
sense, for fixed numbers of candidates manipulation nicely
reduces---not just regarding ranked pairs but in fact as a general
matter---to control by adding voters, simply by putting into the set
of potential voters to add enough copies of every possible vote.
Unfortunately, this approach doesn't quite work, due to the standard
definition of control by adding candidates bounding rather than
exactly setting the number of additions, which is what one needs as
one's reduction's target to make this connection work.  However, this
worry is easy enough to smooth over that we still can prove the
manipulation case here as a quick consequence of groundwork done for
control.

\begin{theorem}\label{t:manip-param-c}
  For ranked pairs (with any feasible tie-breaking
  functions), 
 manipulation
 is in FPT with respect to the
  number of candidates,
  in both the succinct and nonsuccinct input models,
  for both constructive and destructive manipulation,
  and in both the nonunique-winner model and the
  unique-winner model.
\end{theorem}

\begin{proofs}
We will handle together all the cases of this theorem.
Let us again assume without loss of generality that the candidates 
in the manipulation problem are $1,\ldots,j$ with the distinguished candidate
being 1. The manipulation problem has as its input a set of nonmanipulative votes 
$V$ (or for the succinct version, a list of which types of votes occur at least 
once, along with the multiplicities of each) and the set of
manipulators $W$.

We give an fpt-reduction---for readers not 
familiar with their definition, it
can be found near the start of \sectionref{ss:bound-param}---from 
this
problem to a slightly modified version of control by adding voters.
The difference is that instead of having a bound on
the number of votes which we can add, we will need 
an exact number of votes to be added. 
This modification will change only
one of the constraints in our ILPFP (the constraint 
labeled~\ref{eq:add-bound} in the proof of Theorem~\ref{t:adv-param-c-a})
from being an inequality to an equality, and so it is clear 
that the thus-altered control by adding voters problems 
remain in FPT (by the thus-altered version of 
Theorem~\ref{t:adv-param-c-a} and its proof).
So an fpt-reduction to that problem will establish 
our desired FPT result for manipulation.

Our (modified) control instance will have the same set of candidates, 
and the same distinguished candidate.
The set of initial votes in the control instance will be identical to 
the set of nonmanipulative votes in the manipulation instance.
As for the additional vote set in the control instance, 
it will include $\|W\|$ copies of each 
possible vote (out of all $j!$ votes)
(or for the succinct version, a list of all 
$j!$ vote types, each with multiplicity of $\|W\|$).
Finally, the ``exact'' number of votes to be added 
is equal to $\|W\|$.
This completes the specification of the reduction.

Clearly, a manipulation instance is a Yes instance of 
manipulation 
if and only if
this construction map to a Yes instance of the modified 
control problem.
Furthermore, the mapped-to control instance will have the same
number of candidates as the mapped-from manipulation instance. 
Finally, the reduction 
runs in time $|x|(j!)$, where $|x|$ is the manipulation problem's 
input size. Thus the reduction meets the requirement of an fpt-reduction.
This proves that manipulation is in FPT with respect 
to the number of candidates.
\end{proofs}

\section{Other Results for the Unweighted Case}\label{s:other-results}
Section~\ref{s:weighted} will discuss the weighted case, but we 
first present some additional results regarding the unweighted case.

\subsection{Candidate Control Parameterized on the Number of
  Candidates}
\label{s:brute-force-candidates}
Under our primary parameterization of interest, parameterizing on the
number of candidates, and when considering manipulation, bribery, and
voter control, we achieved FPT results 
using the 
looping-over-frameworks technique, and thus involving 
Lenstra's algorithm.  
In contrast, when considering
candidate control problems parameterized on the number of candidates,
we need not use such a powerful technique.  Instead it is sufficient
to brute-force search over all possible control solutions to see if any
of them are successful.  At every possible value for the parameter,
there are only a constant number of possible solutions to any of the
candidate control problems, and checking the success of the possible
solution will require only
a simple polynomial-time task. 
Thus we have algorithms that 
at each fixed parameter value will have a running time that is 
a (large, parameter-value-dependent) constant
times a small uniform polynomial.  This puts these problems in
FPT\@.  
We mention that for the adding/unlimited adding of 
candidates cases, the parameter can even be taken 
to just be the size of the pool of 
potential additional candidates, regardless of how many original 
candidates there were.

\begin{theorem}\label{t:cand-control-bounded-cand}
  For Schulze elections and for (with any feasible tie-breaking
  functions) ranked pairs, control is in FPT with respect to the
  number of candidates, in both the succinct and nonsuccinct input models,
  for both constructive and destructive control, for all standard
  types of candidate control (adding/unlimited adding/deleting candidates and, in both the
  ties-eliminate and ties-promote first-round promotion models,
  partition and runoff partition of candidates), in both the nonunique-winner model and the
  unique-winner model.
\end{theorem}

\begin{proofs}
In the case of adding candidates, at most all the $2^{||D||}$ possible
subsets of the auxiliary candidate set $D$ need be considered.  In the
case of deleting candidates, at most all subsets of $C$ that 
contain the distinguished candidate $p$ need be considered, and so we 
need look at at most 
$2^{||C||-1}$ subsets.
(In the destructive case,
the definition of this control type forbids
trivially satisfying the goal by deleting $p$.  
In the constructive case, deleting $p$ would make success impossible
and so it need not be considered.)
In the
case of runoff partition, there are $2^{||C||-1}$ interestingly 
distinct partitions,
while in the partition case there are $2^{||C||}$.  The difference
between these two 
different types of partition cases is because in runoff partition 
of candidates case the two
parts of the partition are handled symmetrically, and so the partitions 
$(A,B)$ and $(B,A)$ are not interestingly distinct from each other,
and we need consider just one among them.
However, in the partition of candidates case, 
where one side of the partition is 
getting a bye, no such general symmetry can be claimed.
For each of these cases, 
for each setting of its item that we are cycling through above,
we have to call the voting system's winner problem between
one and three times.
So all these cases will be
in FPT for any voting system with a polynomial-time winner 
problem.~\end{proofs}
\subsection{Voter Control Parameterized on the Number of Voters}

Although we feel that the number of candidates is the most natural
parameterization for manipulative action problems,
it is natural to ask about parameterizing on the number of
voters.  We do not exhaustively handle this case for all manipulative
action problems, but we note that voter control problems parameterized
on the number of voters can be shown to be in FPT through simple brute-force.  
Again, as in the case of candidate control problems
parameterized on the number of candidates, we note that the number of
possible solutions to these problems is bounded by a constant for each
parameter value, and checking each solution is easily done in polynomial
time, giving us an FPT algorithm for each of the voter control
problems.  
We mention that for the adding of 
voters cases, the parameter can even be taken 
to just be the size of the pool of 
potential additional voters, regardless of how many original 
voters there were.

\begin{theorem}\label{t:control-voter-as-parameter}
  For Schulze elections and for (with any feasible tie-breaking
  functions) ranked pairs, control is in FPT with respect to the
  number of voters, in both the succinct and nonsuccinct input models,
  for both constructive and destructive control, for all standard
  types of voter control (adding/deleting voters and, in both the
  ties-eliminate and ties-promote first-round promotion models,
  partition of voters), in both the nonunique-winner model and the
  unique-winner model.
\end{theorem}

\begin{proofs}
In the case of adding voters we will have to try at most all of the
$2^{||W||}$ possible subsets of the auxiliary voter set $W$.  In the case
of deleting voters we will have to consider at most all the
$2^{||V||}$ subsets of the voter set $V$.  And in the partition of voters
cases we will have to consider the $2^{||V||-1}$ 
interestingly distinct partitions
of the voter set (again, we need consider just 
one among the partitions 
$(A,B)$ and $(B,A)$).  In all of these cases the large exponential term of
the complexity will be constant with fixed parameter values.  And
beyond that term, we will just have to perform a few iterations of the
voting system's winner function along with a few other simple checks,
putting all these cases in FPT for any voting system with a
polynomial-time winner problem.~\end{proofs}

\subsection{WMG Edge Bound Parameterization}

Let us return to considering the case of parameterization by number of
candidates.  What drove us to our approach of looping over the winner
``frameworks'' we defined, rather than just looping over all WMGs?
It was the fact that even for fixed numbers of candidates, the number
of WMGs blows up as the number of voters increases.  We mention in
passing, though, that if one, \emph{in addition} to parameterizing on
the number of candidates, requires that the absolute value of the edge
weights in the WMG be bounded by some fixed constant independent of
the number of voters, then for that particular special case, one could
loop over all WMGs.  Is this natural and important?  We would tend to
say ``no,'' because assuming that all edges in the WMG have weights
bounded by $k$ is to assume that even as the number of voters grows,
every single head-on-head contest between pairs of candidates is very
evenly matched.  That simply is not the case in most natural
elections.  

On the other hand, perhaps surprisingly, there is something
theoretical to be 
gained from
the strange approach just mentioned of considering elections in
which all edges of the WMG turn out to have relatively low weights.
In particular, we observe that in the NP-hardness-establishing
reductions \emph{to} Schulze control problems used by Menton and
Singh~\cite{men-sin:c:schulze}, all edges in the WMG 
have absolute
value at most 6 (and for some types of control, at most 4 or 2).  
That, along with that fact that all weights of edges in the WMG
have the same parity, gives us 
Corollaries~\ref{c:men-sin} and~\ref{c:men-sin-2}.

\begin{corollary}[{\rm Corollary to the proofs of Menton and Singh~\cite{men-sin:t4ShowVersion-OnlyRevisedDate-may-2013:schulze}}]\label{c:men-sin}
Even when
  restricted to instances having all pairwise contests so equal
  that each WMG 
edge\footnote{In 
Corollaries~\ref{c:men-sin} and~\ref{c:men-sin-2}, when we speak of 
restricting a WMG, the WMG we are speaking of as having to obey
the restriction is 
the WMG
involving \emph{all} candidates
  involved in the problem.  So for adding candidates, the WMG this is
  speaking of is,
  using the votes in the problem instance, the WMG 
  whose nodes are all the 
  initial candidates and all the candidates in the pool of candidates
  that can potentially be added.  Since membership in 
  NP for all the problems discussed is obvious, what is most interesting
  in this theorem is NP-hardness.  And the fact
  that our restriction is applying to the broadest WMG involved in these
  control-by-candidates problems makes the 
  restriction harsher than if it were applying to some sub-WMG, and so
  makes the results stronger.}
 has 
  absolute value at most 2, Schulze elections are NP-complete (in the nonunique-winner
  model) for constructive control by deleting candidates.
  The same holds for constructive control by adding candidates,
  unlimited adding of candidates, 
  partition of 
  candidates in the ties-eliminate model, and
  runoff partition of 
  candidates in the ties-eliminate model,
  except with a bound of 4.
  The same holds for constructive control by
  partition of 
  candidates in the ties-promote model and
  runoff partition of 
  candidates in the ties-promote model,
  except with a bound of 6.
\end{corollary}

Clearly, this immediately implies the following result 
(still keeping in mind
that the values of all edge weights are of the same parity).

\begin{corollary}[{\rm Corollary to the proofs of Menton and Singh~\cite{men-sin:t4ShowVersion-OnlyRevisedDate-may-2013:schulze}}]\label{c:men-sin-2}
  Even when
  restricted to instances having all pairwise contests so equal
  that the total cardinality of the set of absolute values 
  of WMG edges (in the WMG involving all candidates in the instance) 
  is at most 2, 
  Schulze elections are NP-complete (in the nonunique-winner
  model) for constructive control by deleting candidates. 
  Even when
  restricted to instances having all pairwise contests so equal
  that 
  the total
  cardinality of the set of values of WMG edges is at most
  3, Schulze elections are NP-complete (in the nonunique-winner
  model) for constructive control by deleting candidates. 
  The same claims hold for constructive control by adding candidates,
  unlimited adding of candidates, 
  partition of 
  candidates in the ties-eliminate model, and
  runoff partition of 
  candidates in the ties-eliminate model,
  except with the 2 and 3 above replaced by 3 and 5.
  The same holds for constructive control by
  partition of 
  candidates in the ties-promote model and
  runoff partition of 
  candidates in the ties-promote model,
  except with the 2 and 3 above replaced by 4 and 7.
\end{corollary}

\subsection{Adding/Deleting Bound Parameterization}
\label{ss:bound-param}

For those control problems having 
as part of their inputs a limit on how many candidates or voters 
can be added/deleted, it is natural to consider parameterizing 
on that limit.
This parameterization has been studied in 
some voting systems and the relevant problems have often been
found to be W[1]-hard or W[2]-hard, and thus very unlikely to be
fixed-parameter tractable.  For example, 
under
this parameterization,
Betzler and 
Uhlmann~\cite{bet-uhl:j:parameterized-complecity-candidate-control} showed,
for what are known
as Copeland$^{\alpha}$ elections, that 
constructive control by adding candidates and
constructive control by deleting candidates are W[2]-complete,
Liu 
and Zhu~\cite{liu-zhu:j:maximin} proved, for maximin elections,
that constructive
control by adding candidates is W[2]-hard, and Liu and Zhu also achieved
W[1]-hardness results for the relevant voter control problems. For additional
control results parameterized on the problem's internal addition/deletion
limit,
see Table~8 
of Betzler et al.~\cite{bet-bre-che-nie:c:parameterized-elections-survey}.

Hardness for these classes is defined in terms of fpt-reductions
(by which we will always mean many-one fpt-reductions).
Thus one typically shows a problem is, for instance, W[2]-hard by
providing such a reduction from a known W[2]-hard problem.  
We now
 give the standard definition 
of 
fpt-reductions for the case of 
reductions from a problem, 
call it $Q$, with respect to a parameter $j$
(let $j(x)$ be the function that given 
an input to 
$Q$ gives
the value of that parameter on input $x$), to a problem,
call it $Q'$, with respect to a parameter $j'$
(let $j'(x)$ be the function that given 
an input to 
$Q'$ gives
the value of that parameter on input 
$x$).  A function $R$ is 
an fpt-reduction from $Q$ to $Q'$ if the following three conditions
hold~\cite{flu-gro:b:parameterized-complexity}:
(i)~For each $x$, $x\in Q$ if and only if $R(x) \in Q'$.
(ii)~There exist a polynomial $p$ 
and a computable function $f$ such that 
$R$ is computable 
in time $f(j(x)) p(|x|)$.
(iii)~There exists a computable 
function $g$ such that, for all $x$, $j'(R(x)) \leq g(j(x))$.

We observe that, with respect to parameterizing 
on the internal addition/deletion bound, 
for Schulze elections it holds that  
constructive control by adding voters, constructive
control by deleting voters, and constructive 
control by adding candidates are all
W[2]-hard.
\begin{theorem}\label{t:w-2-and-two-others}
Parameterized on the adding
bound, each of 
\begin{enumerate}
\item
constructive control by adding voters,
\item constructive
control by deleting voters, and 
\item 
\label{i:w2p3}
constructive control by adding candidates 
\end{enumerate}
is W[2]-hard for Schulze elections,
in the nonunique-winner model.
\end{theorem}
These problems have already been shown to
be NP-complete by
Parkes and Xia~\cite{par-xia:c:ranked-pairs}
and Menton and Singh~\cite{men-sin:t1ShowVersion:schulze}
through reductions from exact cover by 3-sets (X3C).
In order to prove W[2]-hardness, we need to 
reduce from a parameterized problem that is known
to be W[2]-hard. One such problem is hitting set.
Thus we modify the NP-hardness
proofs given by Parkes and Xia to reduce 
from hitting set instead of from X3C\@.
All that is necessary is to replace 
the role of candidates modeling X3C 
elements with the role of candidates
modeling hitting set sets, and to replace the role of voters 
modeling the
X3C sets with the role of voters with the same preferences 
modeling hitting set elements.
For the sake of completeness, 
we give the complete reduction for one of the 
above three cases, namely, control 
by adding candidates. First, we give the definition
of hitting set.

%
%
%
%
%
%
%
%

%
%
  %
  %
  %
%
%
%
%
%
%
%
%
%
%
%
 %
%
%
%
%
%
%
%
%
%
%
%
%
%
%
%
%

%
%
%

%
%

%

%
%

%

%

%

%

%
%
%
%
%
%
%
%
%
%
%
%
%
%
%
%
%

%

%

%
%
%
%
%

%

%

%

%
%
%
%

%

%

%

%

%
%
%

%

\begin{definition}[Hitting Set]\label{d:hitting}
Given a set of elements $U$, a 
collection  $\mathcal{F}$ of subsets of
$U$, and 
a positive integer $k$, does there exist $H \subseteq U$, with $\|H\| \leq k$,
such that for every $S \in \mathcal{F}$, we have $S \cap H \neq \emptyset$ (i.e., $H$ hits 
every set in $\mathcal{F}$)? 
\end{definition}

\begin{proofsof}{of part~\ref{i:w2p3} of Theorem~\ref{t:w-2-and-two-others}}
The standard NP-complete problem hitting set is also known to be 
W[2]-complete (see p.~464 of~\cite{dow-fel:b:parameterized}).
We give an fpt-reduction from hitting set to our problem.

Given a hitting set instance $(U, \mathcal{F}, k)$ as described in
the definition, 
we will construct a control
instance $(C,D,V,p,k)$ where $C$ is a set of original candidates, 
$D$ is a set of auxiliary candidates,
$V$ is a set of voters,
$p$ is a distinguished candidate,
and $k$ is an adding bound.
The original candidate set $C$ will contain
the following candidates.

\begin{itemize}
\item The distinguished candidate $p$.
\item A candidate $S$ for every $S \in \mathcal{F}$.
\end{itemize}
The auxiliary candidate set $D$ will contain the following:

\begin{itemize}
\item A candidate $u$ for every $u \in U$.  
\end{itemize}

The voter set $V$ will be as follows.  We will not explicitly
construct the entire voter set, but rather we will specify the weight of 
the WMG edges between the candidates and let the voter set be as
constructed by McGarvey's 
method~\cite{mcg:j:election-graph}.
For readers not familiar with McGarvey's method,
and especially since in the proof 
of Theorem~\ref{ranked-pairs-hard} we use McGarvey's
method in a \emph{weighted-voting} setting, and the method's 
mechanics 
are very slightly
different in such a setting, we
provide as Appendix~\ref{sec:McGarvey} a self-contained 
presentation of McGarvey's method for both the unweighted and the weighted 
cases.

\begin{itemize}
\item For every $S \in \mathcal{F}$, $\netadv(S,p)=2$ 
(and so 
$\netadv(p,S)= -2$).
\item For every $u \in U$, $\netadv(p,u)=2$
(and so 
$\netadv(u,p)= -2$).
\item For every $u \in U$, and for every $S \in \mathcal{F}$ such that $u \in
  S$, $\netadv(u,S)=2$
(and so 
$\netadv(S,u)= -2$).

\item All WMG edges not set above will be of weight 0.  
\end{itemize}

The distinguished candidate will be $p$ and the adding limit will be
the same limit $k$ as in the hitting set instance.  This completes the
specification of the reduction.
Note that initially---i.e., 
before any adding of candidates---the winners are exactly
the ``$S$'' candidates (unless $\mathcal{F} = \emptyset$, in which 
case $p$ is initially a winner in the control problem, and the hitting 
set instance is a positive instance, so in the case we are already done).

If we map from a positive hitting set instance, we claim that 
we will have a
positive instance of the control problem.  
Why?  Let $H \subseteq U $, $\|H\|
\leq k$, be a solution to the hitting set instance.  We will show that the set of
candidates $D'$ corresponding to the elements from $H$ will be a
solution to the control instance.  First, $\|D'\| \leq k$, so we are
within the adding bound.  Also, since the hitting set solution
includes members of every set in $\mathcal{F}$, there will be a path
of 
strength two from
$p$ to each candidate corresponding to those sets, as including a
candidate $u \in U$ creates paths from $p$ to every 
``$S$'' candidate hit by $u$. 
It is easy to see that 
$p$ will have paths  to
every other candidate just as 
strong as they have back to $p$, 
and so $p$ will be a
Schulze winner.  

If our reduction maps to a successful control instance, 
we claim it must have mapped from a positive
hitting set instance.  Why?
Recall that we earlier handled the case $\|\mathcal{F}\| = 0$.
Initially, each ``$S$'' candidate will have a path of strength two 
to $p$, but $p$'s strongest path to each of them is of strength 
negative two.
Adding ``$u$'' candidates will leave those strength-two paths from 
each ``$S$'' candidate to $p$ intact. Since all edges have weight 
at most two, regardless of how many candidates we add, no path can have 
a strength of more than two.  So for a control instance to succeed it 
must give $p$ a strength-two path to each ``$S$'' candidate.  In 
our setting, that means adding a set of $k$ ``$u$'' (auxiliary) candidates 
such that each ``$S$'' candidate is pointed to (with a weight-two edge) by at 
least one of them.  However, given our current setup, that itself
says that the hitting set instance being mapped from is a positive instance.

Note also that our reduction clearly runs in polynomial time in
the size of the entire input, easily meeting 
the running-time limit for an
fpt-reduction.  
Additionally, the parameter in the mapped-to instance will always be
bounded by---in fact, identical to---the parameter in the mapped-from
instance. 

So this clearly is an fpt-reduction from hitting set to our problem.
Thus we have established that constructive control by adding
candidates, parameterized on the adding bound, is W[2]-hard for Schulze
elections,
in the nonunique-winner model.~\end{proofsof}

\section{Weighted Case}
\label{s:weighted}
Finally,
although in this paper we have focused on manipulation
problems without weights, and on bribery problems without weights or
prices, we mention 
in passing 
that (keep in mind we still are also
parameterizing on the number of candidates, and that when weights and
prices are used in problems, they are typically taken to 
be---and here we do take them to be---nonnegative 
integers) if one parameterizes by also bounding the
maximum voter weight (if there are weights) and the maximum voter
price (if there are prices), our main theorems 
hold even
in the context of weights and prices.  That is because when weights and
prices are bounded, one can clearly still carry out the approach
we use.

In fact, we can go slightly further, although at the outer edge of
things doing so will require some surgery on our approach.  Not just for
the cases of bounded weights and prices, but even for the case where
there is a bound on the \emph{cardinality} of the set of weights (if
there are weights) and there is a bound on the \emph{cardinality} of
the set of prices (if there are prices), all theorems
(again, still parameterizing also on number of candidates) 
of Section~\ref{s:loops} in bribery and manipulation
still hold (as also do our 
Section~\ref{s:loops} theorems on control, if one 
looks at control in the context of weighted votes;
studying control in the context of weighted votes has only very recently 
been generally 
proposed~\cite{fal-hem-hem:c:weighted-control},
see also~\cite{rus:thesis:borda,lin:thesis:elections}).
To give an example of how we can handle this, we now state and prove in
detail the 
case of bribery 
(We note that both Theorem~\ref{t:bribery-schulze} 
and~\ref{t:bribery-ranked-pairs} follow from
Theorem~\ref{t:section-7-1}, which is not surprising as 
Theorem~\ref{t:section-7-1}
in fact is building on and generalizing their
approach.  We included those earlier 
proofs because they provide a clearer, simpler
setting in which to present our proofs, and they prepare us 
for the proof of the present more flexible case.)
\begin{theorem}\label{t:section-7-1}
  For Schulze elections and for (with any feasible tie-breaking
  functions) ranked pairs, 
 weighted bribery, priced bribery, and weighted, priced bribery
 each are in FPT with respect to the combined parameter ``number of
 candidates'' and ``cardinality of the set of voter weights'' (for
 cases with weights) and
 ``cardinality of the set of voter prices'' (for cases with prices),
  in both the succinct and nonsuccinct input models,
  for both constructive and destructive bribery,
  and in both the nonunique-winner model and the
  unique-winner model.
\end{theorem}

\begin{proofs}
  We will describe how to handle the weighted, priced case.
  This is a generalization of both the weighted, unpriced case 
  and the unweighted, priced case.  So our algorithm will also handle
  both those cases.  The top-level loop will be as in the standard
  bribery proof in Section~\ref{ss:bribery}.  We will specify the
  extensions to the ILPFP construction that are necessary to handle
  the weights and prices.

Let $1, \dots, j$ be the candidates, let $w = \{w_1, \dots, w_y\}$ be the 
set of
weights of the voters, and let $p = \{p_1, \dots, p_z\}$ be the
set of prices.  We will have a constant
$n_{i}^{\alpha,\beta}$ denoting how many voters there are of type $i$ with weight $w_\alpha$ and
with price $p_\beta$ for every $i$, $1 \leq i \leq j!$, $\alpha$, $1 \leq
\alpha \leq y$, and $\beta$, $1 \leq \beta \leq z$.  We will have a
variable $m_{i, \ell}^{\alpha,\beta}$, for every $i$, $1 \leq i \leq j!$,
$\ell$, $1 \leq \ell \leq
j!$, $\alpha$, $1 \leq \alpha \leq y$, and $\beta$, $1 \leq \beta \leq z$.  $m_{i, \ell}^{\alpha,\beta}$
describes the number of voters with weight $w_\alpha$ and price $p_\beta$ that
are bribed from vote type $i$ to vote type $\ell$.  Our total number of
variables is larger than in the unweighted unpriced
case, but it is still bounded as a function of only 
$j$, $y$, and $z$, which make up our combined parameter.

Now we will describe how to build the constraints in our ILPFP\@.  As
before, we can implement the predicates $\bigger$ and
$\strictlybigger$---and for ranked
pairs the additional constraints to handle ranked
pairs---in terms of $\netadv(a,b)$, the weight of the edge from
$a$ to $b$ in the WMG\@.  So we 
need only specify how to formulate
$\netadv(a,b)$ and how to 
formulate appropriate housekeeping 
constraints.  
As to $\netadv(a,b)$,
we need to take into account both voters that are bribed
away from 
preferring $a$ over $b$ and voters that are
bribed 
into preferring $a$ over $b$.  And we need to keep track of
 voters of every type, weight, and price.  Still, what we need is very
 much in line with our simpler implementation of this function in the
 standard bribery case.  So our new formulation of $\netadv(a,b)$ will
 to capture it with the following expression:
\begin{multline*}
\sum_{1 \leq \alpha \leq  y}
\sum_{1 \leq \beta \leq z}
\sum_{i \in pref(a,b)} 
 w_\alpha
\left(
n_{i}^{\alpha,\beta}
- 
\sum_{1 \leq \ell \leq j!,~\ell \neq i} 
\left(
m_{i, \ell}^{\alpha, \beta}
\right)
+
\sum_{1 \leq \ell \leq j!,~ \ell \neq i} 
\left(
m_{\ell,i}^{\alpha, \beta}
\right)
\right) \\
-
\sum_{1 \leq \alpha \leq  y}
\sum_{1 \leq \beta \leq z}
\sum_{i \in pref(b,a)} 
 w_\alpha
\left(
n_{i}^{\alpha,\beta}
- 
\sum_{1 \leq \ell \leq j!,~\ell \neq i} 
\left(
m_{i, \ell}^{\alpha, \beta}
\right)
+
\sum_{1 \leq \ell \leq j!,~\ell \neq i} 
\left(
m_{\ell,i}^{\alpha, \beta}
\right)
\right).
\end{multline*}
We have not excluded the unneeded and rather unhelpful
case $m^{\alpha,\beta}_{i,i}\neq 0$.  Also, the four ``$\ell{\neq}i$''s
above could be dropped as they cancel each other.  However, we include
them as they are the natural way to express what is being modeled.

Additionally we need to implement  constraints to make sure
that the bribery action we select is a legal one. Thus, besides the 
natural constraints $m_{i,\ell}^{\alpha,\beta} \geq 0$, ensuring that we 
do not bribe a negative number of voters of a certain type, we have to enforce 
that we do not bribe
more voters than exist of every type.  Thus for every 
$i$, $1 \leq i \leq j!$, $\alpha$, $1 \leq \alpha \leq y$, and $\beta$, $1 \leq \beta \leq z$, we
have the following constraint:

\[
n_{i}^{\alpha,\beta} \geq \sum_{1 \leq \ell \leq j!} m_{i, \ell}^{\alpha, \beta}.
\]

And also we must restrict the total cost of all our bribes to the
bribe-cost limit $k$.  Thus we have the following constraint:

\[
k \geq 
\sum_{1 \leq \alpha \leq  y}
\sum_{1 \leq \beta \leq z}
\sum_{1 \leq i \leq j!} 
\sum_{1 \leq \ell \leq j!} 
p_\beta
m_{i,\ell}^{\alpha,\beta}.
\]

These additional constraints will enforce a legal selection of voters
to bribe.  These, together with our specification
of $\netadv(a,b)$, complete the specification of the WCF-enforcing ILPFP\@.
Overall we have a structure that will have a constant number of
constraints in terms of the parameters, and that will have a solution if
and only if there is a successful bribery action.  Also there will be only a
constant number of 
WCFs
for every fixed set of parameter
values;  in fact it will be constant in terms of just the number of
candidates, and there will be no more WCFs than in the unweighted,
unpriced case.  

Thus we have established that the weighted variant, the priced variant, and 
the weighted, priced variant of bribery
are each in FPT when parameterized on the 
combined parameter
``number of
 candidates'' and ``cardinality of the set of voter weights''
(if there are weights) and
 ``cardinality of the set of voter prices'' (if there are 
prices).~\end{proofs}

Pushing beyond this, for the case of manipulation (still
parameterized by the number of candidates), we can handle even the
case of there not being any bound on the cardinality of the weight set
of all the voters, but rather there simply being a bound on the
cardinality of the set of weights over all \emph{manipulative} voters.

\begin{theorem}
\label{t:w-manip-card}
For (with any feasible tie-breaking
  functions) ranked pairs, it holds that 
 weighted coalitional manipulation
 is in FPT with respect to the combined parameter 
  ``number of candidates'' and ``cardinality of the manipulators' weight set''
  in both the succinct and nonsuccinct input models,
  for both constructive and destructive manipulation,
  and in both the nonunique-winner model and the
  unique-winner model.  The same claim holds for Schulze elections, except 
  limited to the destructive case (in both the succinct and nonsuccinct
  input models, and in both the nonunique-winner model and the 
  unique winner model) and the unique-winner 
  constructive case (in both the succinct and nonsuccinct input 
  models).\footnote{The nonunique-winner model, 
  constructive case for 
  Schulze elections---which the above theorem is carefully avoiding 
  claiming---also holds.  But as noted in Section~\ref{s:related-work},
  a construction of Gaspers et al.~\cite{gas-kal-nar-wal:c:schulze}
  establishes that 
  the weighted constructive coalitional manipulation problem for 
  Schulze elections is in FPT, with respect to the 
  parameter ``number of candidates,'' in the nonunique-winner model.  
  That is a broader result 
  for nonunique-winner model, weighted constructive coalitional manipulation 
  than would be
  the nonunique-winner model, constructive, Schulze version of 
  Theorem~\ref{t:w-manip-card}, and so it would not make sense to 
  include the nonunique-winner model, 
  constructive Schulze case in the statement of 
  Theorem~\ref{t:w-manip-card}.

  To avoid a potential worry that expert readers might have, we now
  briefly address a rather subtle, technical issue related to why we have
  claimed the destructive Schulze case in
  Theorem~\ref{t:w-manip-card}.  Although Gaspers et
  al.~\cite{gas-kal-nar-wal:c:schulze} for the case of weighted
  manipulation of Schulze elections 
  address only the constructive case (and even
  that, only for the nonunique-winner model), it might be natural to
  think that their constructive result---and we have commented that 
  they provide even an FPT algorithm---will also hold for the destructive
  case and yield there an FPT result stronger than that of
  Theorem~\ref{t:w-manip-card}.  However, as mentioned earlier 
  in this paper, 
  their
  constructive result is crucially based on the lovely fact that in
  the case of the weighted constructive coalitional manipulation problem
  for Schulze elections, if one can make a given candidate a winner
  then there is a set of manipulative votes \emph{in which all
    manipulators vote the same way} and that candidate is selected 
  as a winner.
  This does \emph{not} imply, in any obvious way, that one can make
  the same claim regarding the destructive case---preventing a
  candidate from becoming a winner.  Thinking that such an implication
  holds might be very tempting, based on the (untrue) thought that a
  candidate can be kept from winning exactly if some other candidate
  can be made to win.  This thought, however, is untrue in the
  nonunique-winner model, as winners can coexist.  
  In the unique-winner model, something close to this
  does hold though with a 
  rather substantial 
  twist (see footnote~5 of~\cite{hem-hem-rot:j:destructive-control}).
  But the ``all can vote the same''
  insight used by Gaspers et al.~\cite{gas-kal-nar-wal:c:schulze}
  seems potentially fragile, and is currently known
  only for the constructive,
  nonunique-winner case.  Indeed, 
  Menton and Singh~\cite{men-sin:t4ShowVersion-OnlyRevisedDate-may-2013:schulze} 
state that the 
  analogue fails for the constructive, unique-winner case.
  It thus remains an interesting open
  issue whether the weighted destructive coalitional manipulation
  problem for Schulze elections is in FPT when parameterized only on 
  the number of candidates.  This completes our
  explanation of why Theorem~\ref{t:w-manip-card}'s claim about the
  destructive Schulze case (and similarly, its 
  unique-winner model, constructive Schulze case) 
  does not seem to be subsumed by existing results.}
\end{theorem}

\begin{proofs}
Without loss of generality, 
let the candidates be $1,\ldots,j$, with candidate 1 being the 
distinguished candidate. The input specifies
the nonmanipulators as a collection of votes along with their weights
(with, in the succinct case,
identical-vote identical-weight votes not listed separately
but rather listed by the vote, weight, and a positive binary 
integer giving the number of that-vote that-weight voters)
and it specifies (for the more demanding case, that of 
succinct inputs) the set of
manipulators as a list of pairs, $((w_1, t_1), \dots, (w_s, t_s))$,
with there being $t_i$ 
manipulators having weight $w_i$ (with each $t_i > 0$).  
Our top-level loop will be as follows.
\begin{algorithm}
\caption{Top-level loop for weighted manipulation}
\label{alg:loop-wcm}
\begin{algorithmic}

\STATE \noindent{\bf Start}
\FOR{each $j$-WCF $K$}
\IF{candidate 1 (is/is not) (a winner/a unique winner) according to $K$ and $K$ is an
internally consistent, well-formed $j$-WCF}
\STATE (1) build an ILPFP that checks whether there is an assignment
to the votes of $W$ such that $K$'s winner-set 
certification framework is realized by the set of votes $V \cup W$
\STATE (2)  run that ILPFP and if it can be satisfied then halt and 
accept (note: the satisfying settings will even let us output the precise
manipulation that succeeds)
\ENDIF
\ENDFOR
\STATE declare that the given goal cannot be reached through setting the votes of $W$
\STATE {\bf End}
\end{algorithmic}
\end{algorithm}
\smallskip
\noindent

The ``if'' line at the start of the algorithm should be 
set to ``is'' (``is not'') for the constructive (destructive) case,
and to ``a winner'' (``a unique winner'') for the 
nonunique-winner model (the unique-winner model).
So all we need to do is to specify the ILPFP that we build 
inside the loop, for each given $j$-WCF $K$.

Now, to handle this, we must be careful; since there is no limit on
the overall number of weights, we can't have variables capturing how
many voters of each occurring voter weight there are before
manipulation.  Rather, we in our looping algorithm that is building
the ILPFPs use the power of our algorithm to itself precompute all the
parts of the sums (appearing in the constraints) regarding all the
nonmanipulators---so it is our looping algorithm that is putting in
place constants (of the ILPFP) 
that express the sums of the weights of nonmanipulative
voters that have various properties, in particular, 
it will build the constant $n_{a,b}$
describing the total weight of 
the nonmanipulative voters casting votes that prefer candidate $a$
to candidate $b$, for each pair of candidates $a,b$.  The values
$((w_1, t_1), \dots, (w_s, t_s))$ describing the weights and how many
manipulators there are of each weight will be constants of the ILPFP
as well.  The variables of the ILPFP describe how many manipulators of
each weight get manipulated to each particular vote type:
$m_i^\alpha$ for every $i$, $1 \leq i \leq j!$ and $\alpha$, $1 \leq \alpha \leq
s$, describes how many manipulators of weight $w_\alpha$ are assigned
to vote type $i$.

The WCF can be implemented in terms of the $\bigger$ and $\strictlybigger$
predicates and 
$\netadv(a,b)$
as usual, with 
$\bigger$ and $\strictlybigger$
themselves 
implemented in terms of $\netadv(a,b)$.
In this case we will express $\netadv(a,b)$ as the following:

\[
n_{a,b} + \bigg(\sum_{1 \leq \alpha \leq s} \sum_{i \in \pref(a,b)} 
w_\alpha m_i^\alpha 
\bigg) -
n_{b,a} - \bigg(\sum_{1 \leq \alpha \leq s} \sum_{i \in \pref(b,a)} 
w_\alpha m_i^\alpha 
\bigg).
\]

Besides this we just need a few extra constraints to ensure that the
manipulation chosen is a reasonable one.  We include a constraint $m_i^\alpha \geq 0$ for every
$i$, $1 \leq i
\leq  j!$, and  $\alpha$, $1 \leq \alpha \leq s$, ensuring
that we do not try to 
manipulate to some type a negative number of manipulative voters of some 
weight.
And
we include a constraint $\sum_{1 \leq i \leq j!} m_i^\alpha =
t_\alpha$ for every $\alpha$, $1 \leq \alpha \leq s$,
ensuring that we use the same number of manipulators of every 
weight as are present.

With these additions, we can build an ILPFP to encode our WCF that will
be polynomially bounded in size 
for fixed values of the two
parameters, and there will be only a 
constant 
number of such WCFs with
those fixed parameter values.  
Thus our algorithm specified above will run
in uniformly polynomial time with fixed parameter values, and this problem
is in FPT\@.~\end{proofs}

This still is all a
valid framework for our many-uses-of-Lenstra-based approach.  Note
that for the case of the cardinality of the set of weights being 1,
that gives the case of weighted noncoalitional manipulation
mentioned as an aside by Dorn and
Schlotter~\cite{dor-sch:j:parameterized-swap-bribery}, though here
we're handling even any fixed-constant number of manipulators (since
any fixed-constant number have at most a fixed-constant cardinality of
their weight set), and indeed, even a number of manipulators whose
cardinality isn't bounded but who among them in total have a
fixed-constant cardinality of occurring weights.

Recall that we already were proving our 
unweighted manipulation result (Theorem~\ref{t:manip-param-c})
nearly ``for free,'' by drawing on a related control problem's 
groundwork (and in doing so, we were pointing out an interesting 
connection between manipulation and control).  
The strength of Theorem~\ref{t:w-manip-card} is 
such that it gives us an alternative and even more ``for free''
path to seeing that 
Theorem~\ref{t:manip-param-c} holds:
Theorem~\ref{t:manip-param-c} follows
from Theorem~\ref{t:w-manip-card} simply by setting all the weights 
to~1.

It seems intuitively necessary to bound the 
cardinality of
the manipulator weight set to achieve results like the above.  For
ranked pairs we show that that is in some 
sense necessary: We prove that 
weighted constructive coalitional manipulation (with no bound 
on the number of manipulator weights) is
NP-complete in ranked pairs for each fixed number of candidates
starting at five. Thus there cannot be any algorithm that is
polynomial for any fixed number of (at least five)
candidates,
unless
\pisnp\@.  As a weaker consequence, this will also block the existence
of an FPT algorithm for this problem parameterized solely on the number of
candidates.

\begin{theorem}
\label{ranked-pairs-hard}
For any feasible tie-breaking functions, and 
for each fixed number of candidates, $j$, $j \geq 5$, weighted 
constructive coalitional manipulation is 
NP-complete for ranked pairs (under the given feasible
tie-breaking functions), in both
  the nonunique-winner model and the unique-winner model.
Even without the assumption that the tie-breaking functions 
are feasible, NP-hardness will still hold.
\end{theorem}

\begin{proofs}
Containment in NP is immediately clear if the tie-breaking functions
are feasible.
To prove NP-hardness we will reduce from
a known NP-hard problem, partition. Our construction will not make
any assumptions about the tie-breaking functions. Thus we will establish
NP-hardness even for tie-breaking functions that are not feasible.
As mentioned earlier, the Parkes-Xia~\cite{par-xia:c:ranked-pairs} 
framing of ranked pairs, which we are following in
this paper, 
always selects precisely one 
winner (i.e., is resolute).  
The unique-winner model and the 
nonunique-winner model are in effect the same in 
this setting, and our
result will apply for both the unique-winner model  
and the nonunique-winner model.

We first note that if there is a successful (constructive) 
ranked-pairs 
manipulation, then there is a successful (constructive) ranked-pairs 
manipulation
in which each
manipulator puts $p$ in the top position. We will use this property
in both this proof and the proof of Theorem~\ref{t:tight}.
This fact holds in both the nonunique-winner model and 
the unique-winner model
(which in effect are the same here, see above), 
and in both the weighted and the unweighted
cases (as is clear from footnote~\ref{f:monotonic}'s comments 
about weighted-vs.-unweighted and the fact
that the ``repeatedly applying'' approach of 
the next 
paragraph works fine in both cases).

In order to establish this property, we will first establish a 
property called monotonicity:
If $p$ is the winner of a ranked pairs election under which a voter has 
ranked $p$ immediately below another candidate $a$, then if we modify the election
 by swapping 
in this voter's preference list the position of candidates $p$ and $a$ 
while keeping the rest of
his ordering unchanged, $p$ is still a winner under this modified election.
(The monotonicity of ranked pairs is due to Tideman, but 
for clarity as to the result holding even in our 
flexible-tie-breaking, weighted model/case, and 
to naturally introduce a notion of ``encountering'' that we will
extensively use in this proof and that of Theorem~\ref{t:tight},
and to be self-contained, we 
prove monotonicity here.\footnote{\label{f:monotonic}%
Let us comment on this in a bit more detail.
Several formulations of
ranked pairs 
have been shown to
be monotonic~\cite{tid:j:clones,tid:b:voting}.
The formulation of ranked pairs in \cite{tid:b:voting} 
is quite similar to ours, differing in deriving the tie-breaking
ordering over the pairs from the one over the candidates rather than
allowing other orderings, and in allowing voters to provide
nonstrict rankings rather than requiring strict linear
preferences.
However, we argue that if one looks very
carefully at the monotonicity proof 
given in that
work \cite{tid:b:voting}, it can even be seen
to yield a proof valid for our formulation.
As to the difference in preference models, 
Tideman's proof clearly holds when preferences are required to always be
strict, since that is a subcase of the more flexible vote model used
there. As to why his proof holds for our more flexible
tie-breaking model, note that the argument in his proof relies on
no details of 
the tie-breaking function over the pairs except to deliver a 
strict ordering 
over the unordered pairs of candidates from
the
nonstrict ordering over those pairs based on the weights of the WMG
edges; 
crucially, it can be seen by very careful inspection of Tideman's proof that 
\emph{which} particular such strict ordering the 
tie-breaking function on pairs delivers makes no 
difference to his proof as long as the ordering is consistent
with the WMG-edge-weights-based nonstrict ordering of the pairs.
Thus
Tideman's proof 
would apply just as well even when using our more flexible
tie-breaking function over the pairs, since 
even our functions indeed provide the required
action of unambiguously (and efficiently) 
providing a strict order for 
encountering 
pairs,
and one that does respect what, as just mentioned, it must respect.
(The just-made comments were about the tie-breaking 
function
over pairs.
Our tie-breaking function over the individual 
candidates is in the same model as the one used in that work, and 
so requires no specific discussion here.)
Besides that, although that proof 
is given for the unweighted case, it is clear from the definition of
monotonicity that
if monotonicity holds for the unweighted case,
it also holds for the weighted case.
Thus, taking all of these points together, 
one can see that 
that proof can be argued to apply in our case as well (and 
of course it applies for both
the nonunique-winner model and the unique winner model, since they
coincide, as ranked pairs always selects exactly one winner).  
Nonetheless, to be self-contained, and to naturally introduce 
a notion of ``encountering'' that we will extensively use 
in the rest of the proof of Theorem~\ref{ranked-pairs-hard}
and also in the proof of Theorem~\ref{t:tight},
in our main text we give a 
proof of monotonicity 
(phased to clearly work in our model, which as mentioned above
is more flexible that
Tideman's regarding tie-breaking functions).
}
Our proof of course is deeply indebted to and inspired by Tideman's.)
The property we mentioned earlier will then be derived simply by 
repeatedly applying this 
monotonicity property on all the manipulators until they all rank
$p$ in the top position.

We now prove monotonicity.
Assume now that a voter ranks the initially winning candidate, $p$, 
immediately below some candidate 
$a$. Swapping $p$ and $a$ in this voter's preference list (while keeping 
the rest of the ordering unchanged) will affect the WMG of the election 
only by increasing the weight of the edge
$(p,a)$ by two and decreasing the weight of the edge $(a,p)$ by two.

For the rest of this proof, and also for the proof of
Theorem~\ref{t:tight}, 
we use a notion of 
``encountering,'' defined as follows:
The pair $\{x,y\} $ is said to be 
encountered when the ranked pairs
 winner determination algorithm first considers 
either of the edges $(x,y)$ and $(y,x)$.
Each pair is encountered exactly once (although if when it is 
encountered the relationship between its candidates has already been
transitively set, both edges regarding the pair are simply discarded).
Clearly, the order in which the pairs are encountered
by the algorithm is fixed once
the tie-breaking functions and
the weights of the WMG edges 
are specified.

Recall that the weights of all the WMG edges except
$(a,p)$ and $(p,a)$ 
remain unchanged after our 
modification to the initial election. 
This implies that the relative order 
in which pairs 
are encountered can change 
regarding only $\{a,p\}$'s relation to other pairs.
In particular, if under the modified election 
the order of encountering
pairs differs from the initial election,
it differs 
\emph{only}
in moving forward or moving backward the encountering 
of the pair $\{a,p\}$.

We decompose the rest of the proof of monotonicity into three cases.

\medskip

\noindent\emph{Case 1: The edge $(p,a)$ has positive weight
in the initial WMG, or the edge $(p,a)$ has zero weight in the initial WMG
but $p$ is preferred to $a$ according to the tie-breaking function
among candidates.}
In this case, the shifting of $p$ from immediately
below $a$ to immediately above $a$ in the vote of one voter means that 
the maximum weight (``max''ed of course over the actual 
values, not over the absolute 
values)
among $(p,a)$ and $(a,p)$ in the modified graph will be 
two greater than in the original graph, and will in fact occur on $(p,a)$,
which will now certainly have strictly positive weight.  
And so the order in which pairs are 
encountered will be exactly the same except possibly 
for $\{a,p\}$ shifting to an earlier point in that order.

Now, under the initial WMG (and the tie-breaking functions, but we
will now mostly stop mentioning those except when important, as those
functions don't change between the initial and the modified cases),
$p\succ a$ in the final outcome, as $p$ was the winner.

If $p\succ a$ was set directly at the moment $\{a,p\}$ was encountered in the
initial ordering, then under our modified ordering $p\succ a$ will
certainly also be set directly when $\{a,p\}$ is encountered, since
the
encounter is no later 
(and so no prefix before the encounter, as 
the process unfolds under 
the modified graph, can relate $p$ and $a$ transitively,
since if one did, that same prefix would have existed in the initial
graph and would have set their relation transitively), and $(p,a)$ will have
positive weight when it is encountered. 

If $p\succ a$ was set transitively 
(i.e., before $\{a,p\}$ was encountered)
under the initial WMG's ordering of
encounters,
then since in the modified case $\{a,p\}$ is encountered
in the same place or earlier, if it now still comes after the prefix that
transitively set $p\succ a$ the modified case will still have $p\succ a$ (due to
that identical prefix), and if it
now comes before the prefix that transitively set $p\succ a$ the modified
case will have $p\succ a$ because $(p,a)$ is positive. $p$ will remain the winner.

So under Case 1, $p\succ a$ will still hold under ranked pairs with
respect to the modified WMG\@.
Furthermore, in the modified case,
the pairs that are encountered after $\{p,a\}$,
but were encountered in the initial case before $\{p,a\}$ and were
not discarded, will not be
inconsistent with the order $p\succ a$ as this
order was established in the initial case too, at latest when
$\{p,a\}$ was encountered. Thus, for each such pair, at the moment
when that pair
is encountered in the modified case, the same order established
by its encounter in the initial case either has already
been transitively set by a prefix involving $p\succ a$ or will
be fixed by that encounter. Also, as to the pairs that were encountered in the
initial case before $\{p,a\}$ and were discarded, they will also be
discarded in the modified case due to the identical prefix that
transitively set the order for those pairs in the initial case.
Thus the same orders will be fixed in the modified case as in the
initial case, and the final ordering will be the same. As a
consequence, $p$ will remain the winner with respect to the modified WMG\@.

\medskip

\noindent\emph{Case 2: The edge $(p,a)$ has zero weight
in the initial WMG and 
$a$ is preferred to $p$ according to the tie-breaking function
among candidates.}  
The maximum weight of $(p,a)$ and $(a,p)$ 
in the original WMG was zero, as both were zero,
but the maximum is two in the modified WMG\@.
And so the order in which pairs are 
encountered will be exactly the same except possibly 
for $\{a,p\}$ shifting to an earlier point in that order.

If the relationship between $a$ and $p$ was 
not
set transitively in the initial case,
then 
$a\succ p$ would be the outcome of the encounter (due to the specification
of Case 2), contradicting our assumption that $p\succ a$ in the initial 
case's overall outcome.

So in the initial case, $p$ versus $a$ must have been
set transitively (and thus,
necessarily, before $\{a,p\}$ was encountered), and must have been 
set transitively so as to yield $p\succ a$.  In the modified 
case, we have that $(p,a)$ has positive weight and 
the ordering of encounters is unchanged except for $\{a,p\}$ possibly 
moving earlier.  So since $(p,a)$ is positive, if the relationship between
$p$ and $a$ is not transitively set before $\{a,p\}$ is encountered in 
the modified ordering, then $p\succ a$ gets set at that encounter.  And if 
the relationship between $a$ and $b$ is transitively 
set 
on a prefix of the modified ordering before $\{a,p\}$ is encountered,
then that same prefix must have set $p\succ a$ transitively in 
the initial case,
since that prefix will be identical in both
cases
(because the encounter didn't move forward by the modification),
and so it will be set as $p \succ a$ in our modified case.

So under Case 2, $p\succ a$ will still hold under ranked pairs with
respect to the modified WMG\@. Furthermore, similarly to the
argument we made at the end of Case 1, the same orders will be established for
pairs that in the initial case were encountered before $\{a,p\}$ but
now are encountered after $\{a,p\}$ in the modified case.
Thus the same orders will be fixed in the modified case as in the
initial case, and the final ordering will be the same. As a
consequence, $p$ will remain the winner.

\medskip

\noindent\emph{Case 3: The edge $(p,a)$ has negative weight
in the initial WMG\@.}
Note that, unlike our other two cases, in this case the modification
decreases by two the maximum among the weights of $(p,a)$ and $(a,p)$, and 
so the modified ordering of encounters will be identical to the original
except the pair $\{a,p\}$ may 
be pushed back to a later location in the ordering.

Now, if the relationship between $a$ and $p$ was not set transitively
in the original, then we would have had $a\succ p$ in the original case's
outcome, since $(a,p)$ has positive weight in the original case.  But
we know that $p\succ a$ in the original case's outcome, and so in the
original, $p\succ a$ must be fixed transitively before $\{a,p\}$ is
encountered.

But $\{a,p\}$ is encountered in the same place or later, so whatever
prefix transitively 
set
$p\succ a$ in the original will also occur, and 
transitively sets $p\succ a$, in the modified case, all before $\{a,p\}$ 
is encountered.

So under Case 3, the pair $\{a,p\}$ gets discarded in both the initial
and the modified election, and since we already noted that 
the order of encountering 
pairs does not change by the modification except possibly for
$\{a,p\}$'s relation to other pairs, the orders will be fixed
 in the modified case in exactly the same way
as in the initial case and as a consequence, $p$ will remain the winner.

This concludes our proof of monotonicity for ranked pairs, for both
the weighted and unweighted cases. And as we discussed earlier, we can
repeatedly apply the above-mentioned modification to the preference
list of the manipulators until they all rank $p$ in the top position
(while keeping the rest of the ordering unchanged); so for ranked
pairs, if there is a manipulation making $p$ win, then there is a
manipulation making $p$ win in which all manipulators rank $p$ first.

We now move on to give this proof's reduction, which is from the 
partition problem.  
(Recall that a partition of a set $A$ is a pair 
of disjoint sets $A_1$ and $A_2$ such that $A_1 \cup A_2 = A$.)

\begin{definition}[Partition, see~\cite{gar-joh:b:int}]
Given a list of $n$ integers, $k_1,\ldots,k_n$, does there exist a
partition of $\{1,\ldots,n\}$ into two sets, $I_1$ and $I_2$, 
such that $\sum_{i \in I_1} k_i = \sum_{j \in I_2} k_j$?
\end{definition}

Fix a $j$, $j \geq 5$.
Given a partition instance,
$k_1,\ldots,k_n$, we will construct a weighted 
constructive coalitional manipulation instance 
$(C,V,W,p)$, where $C$ is a candidate set with 
$\|C\| = j$, $V$ is a 
nonmanipulative voter set (each having a weight and voting by a 
tie-free linear order), $W$ is a collection of 
manipulative voters (each starting as a blank-slate vote and having 
a weight), and $p$ is a distinguished candidate.

The candidate set, $C$,
will contain five important candidates: 
the distinguished candidate $p$ and four other
candidates: $a_1$, $a_2$, $b_1$, and $b_2$. There will also be $j-5$ extra candidates
that we will ensure are always easily beaten by $p$.
The extra candidates are added just for the purpose of 
reaching the specified number of candidates.
(Recall that we fixed $j$ in the beginning. Thus we present 
a construction that will directly work for each $j$, $j \geq 5$. 
Alternatively, we could have 
presented the construction for five candidates first, and 
then
extend it to larger numbers of candidates.)

Let $S=\sum_{i=1}^{n} k_i$. 
We will have $n$ manipulators and their weights 
will be $6k_1,6k_2,\ldots,6k_n$.
The nonmanipulators will,
in polynomial time 
using McGarvey's method (this 
claim is justified by
Appendix~\ref{sec:McGarvey}),
be assigned votes 
and weights such that the important candidates induce the
WMG shown in Figure~\ref{f:nonmanip}, and for every extra candidate $e$ 
and each important candidate $c$ (i.e., 
for each $c \in \{a_1,a_2,b_1,b_2,p\}$), $\netadv(c,e)=18S+4$ 
(so that these edges remain 
large---at least $12S+4$---regardless 
of the manipulators' votes).

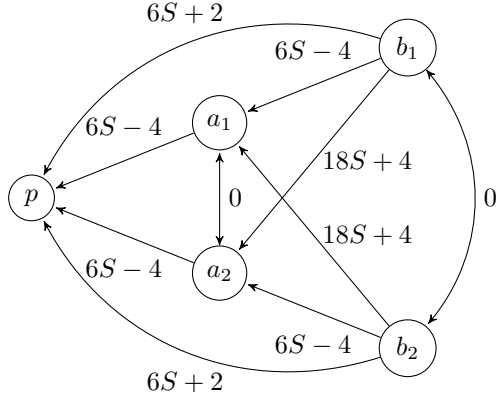
\begin{figure}[tp]
\centering
\begin{tikzpicture}[->,>=stealth',bend angle=40, shorten >=1pt,auto,node distance=4cm]

\tikzstyle{main node}=[shape=circle,draw]
  \node at (0,0) [main node] (p) {$p$};
  \node at (2.5,1) [main node](a1) {$a_1$};
  \node at (2.5,-1) [main node] (a2) {$a_2$};
  \node at (5,2) [main node] (b1) {$b_1$};
  \node at (5,-2) [main node] (b2) {$b_2$};

  \path
    (a1) edge node [above = 0.2 cm] {$6S-4$} (p)
    (a2) edge [<->] node [right] {$0$} (a1)
            edge node [below = 0.2 cm] {$6S-4$} (p)

    (b1) edge [bend right] node[above = 0.15 cm] {$6S+2$} (p)
           edge node[right] {$18S+4$} (a2)
           edge node[above = 0.2 cm] {$6S-4$} (a1)
           edge [<->,bend left] node[right] {$0$} (b2)

    (b2) edge [bend left] node[below = 0.15 cm] {$6S+2$} (p)
          edge node [below = 0.2 cm] {$6S-4$} (a2)
           edge node[right] {$18S+4$} (a1);
\end{tikzpicture}
\caption{The important part of the WMG with just the nonmanipulators having 
cast their votes.  (Most back-edges are left implicit, using the fact that
the weight of an edge $(a,b)$ in a WMG is negative one times the weight
of $(b,a)$.)}
\label{f:nonmanip}
\end{figure}

Now we will show that there is a solution to the partition instance if and only if
there is a solution to the manipulation instance.

We will show that if there is a solution to the mapped-to manipulation
instance, there must be a solution to the mapped-from partition instance.
Suppose there is a solution for the manipulation
instance. Namely, there is a set $V'$ of $n$ votes
with weights as per $W$ such that $p$ wins the election $(C, V
\cup V')$.
Due to the claim established earlier in this proof, we know that if such a $V'$
exists then such a $V'$ exists in which each manipulator ranks $p$
first; so we will take $V'$ to have $p$ at the top of each
manipulator's vote.

Let us use $\netadv'(a,b)$ to denote the weight of the WMG edge $(a,b)$ 
after manipulation.
Thus we will have $\netadv'(p,a_1)=\netadv'(p,a_2)=4$ and $\netadv'(b_1,p)=\netadv'(b_2,p)=2$.
Also the weight of the edges from $p$ to the extra candidates will 
become so large that under ranked pairs their 
relationship with $p$ 
is fixed (namely, $p$ beats each)
before the relationships between $p$ and the important candidates are set.
So, in the final ranking, $p$ is clearly placed higher than 
all extra candidates (we do not
care about the relative order between pairs of extra candidates, since that does not affect
$p$'s performance). Also, we will have $\netadv'(b_1,a_2) \geq 12S+4$ and 
$\netadv'(b_2,a_1) \geq 12S+4$, 
regardless of the manipulators' votes. 
These edges will thus be considered before every other edge in the 
subgraph induced by the important candidates. 
Thus the orders $b_1 \succ a_2$ and
$b_2 \succ a_1$ will 
necessarily 
be fixed in the final ranking  
(and, by the way, we'll have that directly, not transitively). 
We argue that at least one of the edges $(a_1,b_1)$
or $(a_2,b_2)$ must achieve a nonnegative weight.
Otherwise there will not be any path of nonnegative weight edges from $p$ to
$b_1$ and $b_2$, and since $\netadv'(b_1,p) = 2$, that means 
no sequence of candidate pairs that can 
transitively establish
$p \succ b_1$ or $p
\succ b_2$ in the final ranking can possibly all be encountered
come before $(b_1,p)$
and $(b_2,p)$ are considered,
and so 
we will eventually have $b_1 \succ p$ and $b_2
\succ p$ in the final ranking.
But since the manipulator weights are all multiples of 6
(and the total manipulation weight is $6S$ and 
$\netadv(a_1,b_1)=\netadv(a_2,b_2) = (-6S)+ 4$), 
the only nonnegative weight the edges $(a_1,b_1)$ and $(a_2,b_2)$
can possibly achieve is 4. In particular $D'(a_1,b_1) \geq 0$ only when
$D'(a_1,b_1)=4$, and this happens only 
if all the manipulators prefer $a_1$ to $b_1$. Analogously $D'(a_2,b_2) \geq
0$ only when $D'(a_2,b_2)=4$, and this happens only if all the
manipulators
prefer $a_2$ to $b_2$.

We have established that at least one of 
the edges $(a_1,b_1)$
or $(a_2,b_2)$ must achieve a nonnegative weight.
Now, without loss of generality (as the other case---namely
that $(a_2,b_2)$ achieves nonnegative weight---is symmetric),
assume that 
$(a_1,b_1)$ achieves nonnegative weight.  And so, as we have 
argued, it even holds that 
$\netadv'(a_1,b_1)=4$.
We argue 
that $\netadv'(b_1,b_2)=0$. Otherwise, we would have one of two cases
(note that the manipulator weights are all multiples of 6 and
 $\netadv(b_1,b_2) = 0$, so $D'(b_1, b_2)$ must be a multiple
of 6):
\begin{enumerate}
\item{$\netadv'(b_1,b_2) \geq 6$:} In this case, $(b_1,b_2)$ will be considered before $(a_1,b_1)$
and the order $b_1 \succ b_2$ will be fixed in the final ranking. 
Since we already have the order $b_2 \succ a_1$,
by transitivity, we have $b_1 \succ a_1$ and $(a_1,b_1)$ gets discarded. 
Thus
the order between $p$ and $b_1$ will not be specified transitively by
any 
sequence of fixed orders 
before the edge $(b_1,p)$ is considered. 
And when $(b_1,p)$ is considered, $b_1\succ p$ would be set.
So
$p$ would not be a winner, contradicting our assumption that 
we have a successful manipulation on the given instance.
\item{$\netadv'(b_2,b_1) \geq 6$:} We have two subcases here
(keeping in mind still that $\netadv(b_2,a_2) = 6S-4$,
the manipulators total $6S$ in weight, and each manipulator's 
weight is a multiple of 6).
\begin{enumerate}
\item{$\netadv'(a_2,b_2) = 4$:} In this case, $(b_2,b_1)$ 
will be considered before $(a_2,b_2)$ and the order $b_2 \succ a_2$ will 
be fixed by transitivity before considering
that edge. Then the only nonnegative incoming edge to $b_2$, which is $(a_2,b_2)$, 
gets discarded, and eventually, when $(b_2,p)$ is considered, $p$ will be
ranked lower than $b_2$.
\item{$\netadv'(b_2,a_2) \geq 2 $:} In this case, since there will not
  be any nonnegative incoming edge
to $b_2$, $p$ will be ranked lower than $b_2$ when considering $(b_2,p)$.
\end{enumerate} 

\end{enumerate}
Thus it must hold that $\netadv(b_1,b_2)=0$. We now can build a solution for the 
partition problem as follows. For every manipulator $i$ (which has the weight $6k_i$) 
having ranked $b_1$ higher
than $b_2$, we make $I_1$ include $i$, and for every manipulator $j$ (which has the 
weight $6k_j$) having ranked
$b_2$ higher than $b_1$, we make $I_2$ include $j$. Then $(I_1,I_2)$ is 
clearly
a solution to the partition problem.

We will show that if there is a solution to the mapped-from partition
instance, then there must be a solution to the mapped-to manipulation
instance.  Suppose there is a solution for the partition problem. That
is, we have a partition of $\{1,\ldots,n\}$ into $I_1$ and $I_2$,
such that 
$\sum_{i \in I_1} k_i = \sum_{j \in I_2} k_j$.
We make every
manipulator $i$ with $i \in I_1$ cast the following vote:
\[ p > a_1 > a_2 > b_1 > b_2 > \ldots \]
(where $[\ldots]$ denotes the extra candidates in any arbitrary order),
and make every manipulator $j$ with $j \in I_2$ cast the following vote:
\[ p > a_2 > a_1 > b_2 > b_1 > \ldots\;. \]

After the manipulation, we will have $\netadv'(a_1,a_2)=\netadv'(b_1,b_2)=0$ and $\netadv'(a_1,b_1)=\netadv'(a_2,b_2)=4$.
So, after having fixed in the final ranking the orders $b_1 \succ
a_2$,  $b_2 \succ a_1$, $p \succ a_1$, and $p \succ a_2$,  the edges
$(a_1,b_1)$ and $(a_2,b_2)$ will be considered, giving the final orders $p \succ a_1 \succ b_1$ and $p \succ a_2 \succ b_2$.
So, we will have a transitive order from $p$ to both $b_1$ and $b_2$ and $p$ will be a winner.~\end{proofs}

One might naturally wonder whether the constant five is tight for the
NP-hardness result presented in Theorem~\ref{ranked-pairs-hard}. 
Theorem~\ref{t:tight} proves that five indeed is tight here, by 
showing that for each number of 
candidates smaller than five, we can
solve the manipulation problem under consideration
in polynomial time. The key idea is
to show that,
for each ranked-pairs instance with fewer than five candidates, 
it holds that if there exists a successful 
manipulation, 
then 
for that same instance
there
exists a successful manipulation 
in which all the manipulators vote the same. 
\begin{theorem}\label{t:tight}
For any feasible tie-breaking functions, and 
for each number of candidates, $j$, $j < 5$, weighted 
constructive coalitional manipulation is 
solvable in polynomial time for ranked pairs (under the given feasible
tie-breaking functions), in both
  the nonunique-winner model and the unique-winner model.
\end{theorem}

\begin{proofs}
We will show that in instances with fewer than five candidates, 
if a successful manipulation exists, then there
exists a manipulation where all the manipulators cast the same vote.

In light of this fact, 
there are only a constant number
of vote types to check regardless of the number of manipulators. 
Thus we merely need, for each vote type, to run the ranked-pairs
winner determination procedure on the election that consists of the
nonmanipulators having cast their votes, and the manipulators 
all having cast a vote of that particular type. Thus we have a 
polynomial-time algorithm.

All that remains is to 
prove, for every number of
candidates smaller than five, 
that if a successful manipulation exists, then there
exists a manipulation where all the manipulators cast the same vote.
For each number of candidates less than five,
we first assume that there exists
a successful manipulation and then argue that we can still achieve the
goal, even if all the manipulators cast identical votes.
(In fact, we will even argue that if the number of candidates is 
less than five and 
a successful manipulation exists, then there
exists a successful manipulation in 
which all the manipulators cast the same vote
and that vote has $p$ as the most preferred candidate.)
As before our result immediately applies to both winner models, since
ranked pairs always selects exactly one winner.  

Recall that, regardless of the number of candidates, if there exists a
manipulation, then there exists a manipulation where all the
manipulators rank $p$ at the top position
(this claim is
discussed and justified in the proof of 
Theorem~\ref{ranked-pairs-hard}). 
So we can safely 
assume that, in the successful manipulation, all the manipulators rank
the distinguished candidate at the top position. This immediately
proves our claim for $j \leq 2$.  For the remaining 
values of $j$ that we need to handle, namely $j=3$ and $j=4$, we 
will show that 
if we have a successful manipulation, we can make $p$ win with all the manipulators
agreeing on the rest of their rankings in addition to all ranking $p$ first.

For $j=3$, let the set of candidates be $C=\{p,a,b\}$ with $p$
being the distinguished candidate. As we discussed earlier, all the
manipulators rank $p$ at the top position. This lets us know,
beforehand, the
weight of the WMG edges $(p,a)$ and $(p,b)$ after the manipulation,
regardless of the manipulators' preferences over $a$ and $b$.

We now consider the moment when the pair $\{a,b\}$ is 
encountered (the notion of encountering is as defined in the proof 
of Theorem~\ref{ranked-pairs-hard}). At that time, the order between $a$ and $b$ has already
been fixed in the
final ranking only if it has been fixed transitively by a sequence of
orders including
$p$. However, this can happen only if 
$a \succ p$ or $b \succ p$ has been already set at that time, 
which cannot happen since we have assumed  
a successful manipulation.
Thus the order between $a$ and $b$, in the final ranking, will be
determined right at the
time when the pair $\{a,b\}$ is encountered. Let us assume without loss of
generality (as the two cases are symmetric) that the order being fixed
is $a\succ b$. We claim that 
all the manipulators can vote $p> a> b$ with $p$ still winning the
manipulated election. This change of the manipulators' votes will not weaken 
the weight of the edge $(a,b)$ and will not change the weight of the
edges involving $p$.
Thus when $\{a,b\}$ is encountered, the order $a\succ b$ will be fixed in
the final ranking 
as before, except that this pair might be encountered earlier,
before the pairs $\{a,p\}$ or $\{b,p\}$. However, earlier fixing of
the order $a\succ b$ cannot prevent $p$ from winning.

Finally, for $j=4$, let the set of candidates be $C=\{p,a,b,c\}$ with $p$ being
the distinguished candidate. Again we are assuming that each
manipulator puts $p$ in his or her top position.

Let us assume without loss of generality that after the manipulation
has occurred, among the pairs not including $p$, the 
ranked-pairs winner determination algorithm encounters the pairs
in the order $\{a,b\}$, $\{a,c\}$, and $\{b,c\}$.

In a similar fashion to the argument we made for the case of $j=3$, 
at the moments when $\{a,b\}$
and $\{a,c\}$ are encountered, the edges between these pairs would get
discarded by a transitively fixed order only if the sequence of orders
establishing that order
includes $p$, which cannot happen since we have assumed a successful
manipulation. So for the first two pairs that are encountered among
these three pairs, their relative order is fixed right at that time 
and this can be done in four different ways:
\begin{enumerate}
\item The orders  $a\succ b$ and $a\succ c$ are fixed. 
\item The orders $b\succ a$ and $c\succ a$ are fixed.
\item The orders $b\succ a$ and $a\succ c$ are fixed.
\item The orders $a\succ b$ and $c\succ a$ are fixed.
\end{enumerate}

In the first two cases, when the pair $\{b,c\}$ is being encountered,
the relative order of $b$ and $c$ has not been fixed yet. 
Thus it will be 
fixed according to whichever of $(b,c)$ and $(c,b)$ is considered first
and according to the tie-breaking function that handles zero-weight edges 
if these edges are of weight zero.
Since $b$ and $c$ are
symmetric in both cases, let us assume without loss of generality
that $b\succ c$ is what gets fixed. We now claim that all the manipulators can cast the vote 
$p>a>b>c$ in the first case (of our above list of four cases) and $p>b>c>a$ in the second case, 
with $p$ still winning the manipulated election.
To argue for the first case, we note this change will not weaken any of the edges
among $(a,b)$, $(a,c)$, and $(b,c)$. What might be affected instead is
the  order in which the pairs
are encountered, either among the pairs $\{a,b\}$, $\{a,c\}$, and
$\{b,c\}$
 or between these pairs and
the pairs involving $p$ (the pairs not involving $p$ may be encountered
before some of the pairs involving $p$ that they used to be
encountered after). The latter is obviously of no concern if we can
show that for each pair of the candidates not involving $p$, the same
order will be
fixed as before at latest when that pair is encountered.
We now argue that the orders being fixed for the pairs not involving $p$ 
will be exactly the same as before. 
Note that since the weights of the edges $(a,b)$,
$(a,c)$, and $(b,c)$ are not weakened by the modification to the
manipulators' votes, for the first two pairs being
encountered among $\{a,b\}$, $\{a,c\}$, and $\{b,c\}$ 
(please keep in mind the ``without loss of generality'' 
assumptions, two and four paragraphs before the 
present one, as to the order of encountering among these), the encounter
of those pairs establishes the same order as before (i.e. $a\succ b$
for $\{a,b\}$, $a\succ c$ for $\{a,c\}$, and $b\succ c$ for
$\{b,c\}$). Furthermore, when the last pair among these three is encountered, 
no sequence of orders has transitively set the order between the
candidates in that pair. So that encounter will also establish the
same order as before. This proves that the final ordering is the same
as before and
all the manipulators can safely cast the vote $p>a>b>c$ in the first
case. The argument for the second case is analogous.

The second two cases, though, are a bit different since the order
between $b$ and $c$
is set transitively by the orders $b\succ a$ and $a\succ c$ in the third case, 
or by the orders $c\succ a$ and $a\succ b$ in the fourth case. 
We claim that in the third case all the manipulators can vote 
$p>b>a>c$ and $p$ will still win the manipulated election.
Note that this set of manipulative votes might strengthen the weight of the edges $(b,a)$ and $(a,c)$ and
$(b,c)$ and allow their corresponding pairs to be encountered in an arrangement different from
the initial manipulation (plus the pairs not involving $p$ may be encountered
before some of the pairs involving $p$ that they used to be
encountered after, but again this is not important if we can show that for
each pair of the candidates not involving $p$, the same order will be
fixed as before at latest when that pair is encountered).
However, under the modified manipulation,
this will not change the orders being fixed
assuming that the pair $\{b,c\}$ is the last
pair being encountered among the three. But even if $\{b,c\}$
is no longer the last pair, we argue that the final ordering will
be exactly the same. Note that since the weights of the edges $(b,a)$
and $(a,c)$ are not weakened by the modification to the manipulators'
votes, the encounter of the first pair among
$\{a,b\}$ and $\{a,c\}$ establishes the same order as before
(i.e. $b\succ a$ for $\{a,b\}$ and $a\succ c$ for $\{a,c\}$),
regardless of when $\{b,c\}$ is encountered. And the encounter of the
second pair establishes the same order as before, except possibly if
that second pair gets discarded due to the order being transitively set (of
course by a sequence of orders not involving $p$).
So the only way the final ordering
might differ from the initial manipulation is either when the
encounter of the pair $\{b,c\}$ establishes the order $c\succ b$ or
when the second-encountered pair among $\{a,b\}$ and $\{a,c\}$
gets discarded by a sequence of orders including an order between $b$
and $c$. The latter itself could happen only if the encounter of the
pair $\{b,c\}$ establishes the order
$c\succ b$ (so that along
with $b\succ a$ it transitively fixes the order $c\succ a$, or that along
with $a\succ c$ it transitively fixes the order $a\succ b$). However, 
the weight of the edge $(c,b)$ under the initial manipulation
must have been at least as large as its weight under the modified manipulation.
Furthermore, the weights of the edges 
$(b,a)$ and $(a,c)$ under the
initial manipulation have to be at most as large
 as their weights under the modified manipulation.
Thus if under the modified manipulation the order $c\succ b$ is ever
set, $\{b,c\}$ could not have been the
last pair that was encountered among the
three under the initial manipulation, which contradicts our assumption. So 
all the manipulators can cast the vote $p>b>a>c$ and $p$ will still
win.  In the fourth case, the manipulators can analogously vote
$p>c>a>b$ 
and $p$ will continue to win.

This proves that for each number of candidates $j$, $j \leq 4$, if
a successful manipulation exists, then there exists a manipulation 
where all the manipulators vote the same. This in turn proves that
weighted constructive coalitional manipulation is solvable in
polynomial time for each number of candidates $j$, $j<5$.
\end{proofs}

\section{Open Problems}
For all cases of (unweighted) 
bribery, control, and manipulation, including many 
where general-case hardness results exist, this paper 
proves that Schulze elections and ranked-pairs elections are
fixed-parameter tractable with respect to the number of candidates.

In order to concisely represent our unweighted results, we 
list in Figure~\ref{fig:fpt-table} 
\begin{figure}[!tbp]
\begin{tabular}{cc|c|c}
  & & \multicolumn{2}{c}{Voting Rules} \\
  \cline{3-4}
  & & Schulze & Ranked Pairs \\
  \hline
\multicolumn{2}{c|}{Bribery}

& \textbf{FPT} [Thm.~\ref{t:bribery-schulze}], NPC \cite{xia:margin-of-victory} & \textbf{FPT} [Thm.~\ref{t:bribery-ranked-pairs}], NPC \cite{xia:margin-of-victory} \\
  \hline
\multicolumn{2}{c|}{Manipulation}
& P \cite{gas-kal-nar-wal:c:schulze} & \textbf{FPT} [Thm.~\ref{t:manip-param-c}], NPC \cite{con-pro-ros-xia-zuc:c:unweighted-manipulation}\\
  \hline
  \multirow{6}{*}{Control by} 
   & adding cand. & FPT [Thm.~\ref{t:cand-control-bounded-cand}], NPC \cite{par-xia:c:ranked-pairs}  & FPT [Thm.~\ref{t:cand-control-bounded-cand}], NPC \cite{par-xia:c:ranked-pairs} \\
   & deleting cand. & FPT [Thm.~\ref{t:cand-control-bounded-cand}], NPC \cite{men-sin:c:schulze} & FPT [Thm.~\ref{t:cand-control-bounded-cand}], NPC \cite{par-xia:c:ranked-pairs}\\
   & partition of cand. & FPT [Thm.~\ref{t:cand-control-bounded-cand}], NPC \cite{men-sin:c:schulze} & FPT [Thm.~\ref{t:cand-control-bounded-cand}]\\
   & adding voters & \textbf{FPT} [Thm.~\ref{t:adv-param-c-a}], NPC \cite{par-xia:c:ranked-pairs} & \textbf{FPT} [Thm.~\ref{t:adv-param-c-a}], NPC \cite{par-xia:c:ranked-pairs}\\
   & deleting voters & \textbf{FPT} [Thm.~\ref{t:adv-param-c-d}], NPC \cite{par-xia:c:ranked-pairs} & \textbf{FPT} [Thm.~\ref{t:adv-param-c-d}], NPC \cite{par-xia:c:ranked-pairs}\\
   & partition of voters & \textbf{FPT} [Thm.~\ref{t:pv-param-c}], NPC \cite{men-sin:c:schulze} & \textbf{FPT} [Thm.~\ref{t:pv-param-c}]\\ \hline
\end{tabular}
\caption{\label{fig:fpt-table}
Best 
current tractability and intractability results regarding the 
unweighted case of Schulze and ranked pairs, for constructive attacks in the nonsuccinct input model
and for the nonunique-winner model. Our 
FPT results that are achieved by the 
looping-over-framework
approach are shown in boldface.}
\end{figure}
the best currently known tractability and intractability results
regarding the unweighted case.
All the figure's FPT
results hold for 
both the constructive and destructive cases, in both the
succinct and nonsuccinct input models, and for
both the unique-winner and nonunique-winner models.
Due to some of the other NPC cases not having been
explored in the literature, though, 
Figure~\ref{fig:fpt-table} limits itself 
to speak just about the 
constructive case in the nonsuccinct input model and
in the nonunique-winner model.

The most striking remaining open direction regards the weighted 
cases.  
For example, Gaspers et al.~\cite{gas-kal-nar-wal:c:schulze} 
proved that the constructive, nonunique-winner case of 
weighted coalitional manipulation is in FPT\@.
Can their result be extended to the 
constructive, unique-winner case, or the 
destructive, unique-winner case, or the 
destructive, nonunique-winner case?  Can one even provide 
a fixed-parameter tractability result for these three cases?
These all remain open questions, although in Theorem~\ref{t:w-manip-card}
we give  
fixed-parameter tractability results for special cases of all 
three of these issues.  The analogous issues are not open 
for ranked pairs:
Theorem~\ref{ranked-pairs-hard}
shows that unless $\p = \np$, the ranked-pairs analogue of the 
Gaspers et al.\ result cannot hold.  

\section*{Acknowledgments}
A two-page 
extended-abstract 
version of this paper
appeared in
AAMAS~2013~\cite{hem-lav-men:c:schulze-and-ranked-pairs}.
We thank anonymous referees for helpful comments and
suggestions,
including the alternate certification 
framework of 
Appendix~\ref{sec:alternate-swcf}
and the suggestion to obtain
Theorem~\ref{t:manip-param-c} 
as a consequence of
Theorem~\ref{t:adv-param-c-a}.
This work was supported 
in part by
NSF grants 
CCF-0915792, 
CCF-1101479, and 
CCF-1116104.

\bibliographystyle{plain}

\appendix

\section{An Alternative Schulze Winner Certification Framework}
\label{sec:alternate-swcf}
The following discussion
may certainly be skipped,
except by curious readers or readers wanting to get more of a sense 
of the flavors that winner certification frameworks can have.
In particular, the following discussion presents
an alternative Schulze winner certification framework
to the one described above; the above-described
framework is the one we used throughout this 
paper.  The alternative framework we will now describe 
in some ways is very clear (as to information it holds 
and counting how many such structures there are) but in some ways is a bit
subtle (as to the time-cost of seeing whether they really certify 
that a given person is a winner).

Our SWCFs, as just described and as we use throughout the 
paper, 
are very explicitly certifying why a given
candidate is a Schulze winner.  We mention that one could
alternatively loop over a different type of structure that many may
find more easier to understand as a clean, clear structure to loop
over since it is basically a triangular matrix with 3-valued entries,
namely, $\{<,=,>\}$.  
Namely, consider a table that for each pair of
edges-between-candidates says whether the first has a weight greater
than, equal to, or less than the other.  This structure certainly has enough
information to tell whether a given candidate is in the Schulze winner
set.  So we could loop over all such tables making our candidate of
interest a winner (or if in the ``destructive'' case, 
not a winner).  The fact that
some such tables can never be realized since they set up violations of
transitivity---e.g., $n_1 < n_2 < n_3 < n_1$---is not a problem for the way
we'll use our SWCFs; we'll use them in 
ILPFPs
related to attack types, and unrealizable tables
of course can never be realized by any attack actions since they
can't be realized at all, and so they will not harm our algorithms.
In some sense, this attractive 
approach is making the SWCFs more oblique as to 
why they ensure that a given candidate is a winner, but since these
structures 
are simply a big ternary triangular matrix, 
they are very simple structures to loop through 
and are very 
simple structures 
for which to see that their number 
is bounded for each fixed number of candidates.
In particular---and we make no attempt to try to take advantage here 
of the ``negative one factor'' relationship between $(a,b)$ and $(b,a)$
weights---there are certainly at most $3^{(p^2-p)/2}$ such structures,
where $p = 2{\|C\| \choose 2}$.

Since we won't return to a discussion of this alternate approach,
we mention now two issue that are relevant to why it would work
as an approach.  In the ILPFPs we'll build later, we'll loop through
the SWCFs and will have to in the ILPFP capture whether the structure
can be realized.  But the tools we build for the framework we actually
use make it very clear that in our ILPFPs we can ensure that 
a collection of $<$, $=$, and $>$ relationships hold, and so it would be 
easy to modify our ILPFPs to enforce the claims of our ternary 
triangular matrices.  The more subtle and interesting issue
is that in our ILPFPs
we need to tell whether a given SWCF makes a given candidate (or precludes
a given candidate from being) a winner.  For the SWCFs we are using in our 
paper, this is a trivial, immediate check.  
But for the ternary triangular matrix approach, which gives us just 
$<$-vs.-$=$-vs.-$>$ information, it is not even clear we can check winnership
in polynomial time in the size of the matrix, given that there are an 
exponential number of potential simple paths between two points.  
However, the key insight to have that quells that worry 
is that that issue is irrelevant to ensuring membership in the class FPT, which
is the goal of this paper.  Even if determining whether a given 
ternary triangular matrix makes someone a winner takes time exponential
in the size of the matrix (which itself is as noted 
above of size roughly quartic 
in the number of candidates), this is a constant for each fixed number of 
candidates, and so does not in any way interfere with our FPT
claims.  Again, in our paper we are not using the ternary triangular 
approach, and so all our ILPFPs are in terms of our primary 
approach.  But this appendix is here 
to make clear that the ternary triangular 
approach certainly would work, if implemented with the care and cautions 
just mentioned.

\section{McGarvey's Method}
\label{sec:McGarvey}
McGarvey's method~\cite{mcg:j:election-graph}, 
is quite simple.  
However, to be self-contained,
and since in one of the three times we draw on it we are using it 
in a weighted-voting setting
(namely, in the proof of Theorem~\ref{ranked-pairs-hard}), where 
it---see below---works very slightly differently, we now 
provide a description of McGarvey's 
method.
Suppose we are given some WMG,
and we wish to generate a set of votes 
realizing it.  The one-candidate case isn't interesting, so 
we assume that there are at least two candidates.
Since we are speaking of a WMG, for each $(a,b)$, $a \neq b$, 
it must hold that 
the weight on $(a,b)$ is negative one times the weight on 
$(b,a)$.  
Also, either all of the edges are even in weight or all are odd.
In particular,
in the case we are now discussing, the unweighted case,
the parity of each edge weight is the parity of the number of voters;
in the
weighted case, which we will discuss 
later in this appendix, the parity of each edge will be the parity of the 
total vote weight.
Suppose the candidates are $a$, $b$, and $c_1 \ldots,c_z$, $z \geq 0$.
Consider adding the following two votes:
$a>b> c_1 > \cdots > c_z$ and 
$c_z > c_{z-1} > \cdots > c_1 > a > b$.
This vote pair increases by 
two the weight of the edge from $a$ to $b$ in the WMG,
decreases by two the weight of the edge from $b$ to $a$ in the WMG, and leaves
all other edges unchanged in the WMG\@.
Of course, this trick can be used not just on $a$ and $b$, but on any 
pair of distinct candidates.

So in the case of unweighted voting, if the (target) edge weights are even 
we can, for each pair of candidates, simply keep adding 
a pair of votes to shift in the right direction, by two, the weights 
on the edges between them, until the desired values are reached. 
If the (target) edge weights are odd, we can 
initially add one arbitrarily chosen vote, thus
making all edges odd in weight, and then can appropriately add pairs of votes as 
before to bring the weights to the desired values.  
This clearly can be done in time polynomial in the sum of 
(a)~the sum of the absolute values of the edge weights, and (b)~the size
of the problem (i.e., the total size of the candidate names
and the total size of the edge weight descriptions, though the latter is 
in fact handled by~(a)).  Of course, if the edge weights are large,
this could take an exponential amount of time unless 
the edge weights are coded in unary.  However, in this paper,
everywhere where we are using McGarvey's method for unweighted voting,
the absolute values of the edge weights are bounded by a small 
constant, so this is not an issue.  

However, when we invoke McGarvey's method in the proof of 
Theorem~\ref{ranked-pairs-hard}, it is used in a weighted-voting 
context, and for WMGs where the sum of the absolute values of the edge weights
often will be exponential in the problem size.  That is no problem, though,
since in the weighted-voting case, in the even-weight case, 
if the desired values of the WMG edges 
$(e,f)$ and $(f,e)$ are $2w$ and $-2w$, with $w$ a nonnegative 
integer, we simply 
add the two votes mentioned above, using $e$ and $f$ in the roles of $a$ 
and $b$, with each vote being a weight-$w$ vote.  And we do this for 
each pair of distinct candidates.  
In the odd-weight case we can again just initially
add an arbitrary weight-one vote to make the 
weights odd, and 
then for every pair we can add the appropriately chosen pair
of votes (of weights of either both $w$ or both $w+1$, 
depending on the current edge weights set by the one initial vote) 
to bring the edge weights to the desired values of $2w+1$ and $-2w-1$.

So for the weighted-vote 
case, we can clearly generate a realizing set of votes in time polynomial
in the size of 
the problem instance (the sum of the sizes
of all the weights---not the sum of 
their values but 
the sum of the sizes of their binary encodings---plus the sum of 
the sizes of the candidate names).

\section{Control Definitions}\label{sec:control-definitions}
In this section, we provide more formal definitions of the 
benchmark control types.  
Since such
papers
as~\cite{bar-tov-tri:j:control,hem-hem-rot:j:destructive-control,fal-hem-hem-rot:j:llull}
provide a detailed discussion of the motivations of and real-world 
inspirations for the benchmark
control types, we don't repeat the discussion here.

The statements below are taken, sometimes
identically, from
Faliszewski et al.~\cite{fal-hem-hem-rot:j:llull}.  
We first define the voter-control types, since those are more central in
this paper.  Then we define the candidate-control types, which in this
paper appear only in Section~\ref{s:other-results}.  All definitions
below are stated for the nonsuccinct, nonunique-winner model.  
Section~\ref{s:other-results}'s comments specify how to modify these
to define the succinct case and/or the unique-winner case.

\subsection{Voter-Control Definitions}

The adding-voter control problems are the following.

\begin{description}
\item[Name:] 
The 
constructive control by adding voters problem
for $\cale$ elections and the 
destructive control by adding voters problem
for $\cale$ elections.
\item[Given:] A set $C$ of candidates, two 
disjoint 
collections of
voters, $V$ and $W$
  (each voter's vote is a tie-free linear order), 
a distinguished candidate
  $p$, and a nonnegative integer~$k$.
\item[Question (constructive):] Is there
  a subset $Q$, $\|Q\| \leq k$, of voters in $W$ such that the
  voters in $V \cup Q$ jointly elect $p \in C$ as a
  winner according to system~$\electionsystem$?

\item[Question (destructive):] Is there
  a subset $Q$, $\|Q\| \leq k$, of voters in $W$ such that the
  voters in $V \cup Q$ do not elect $p$ as a winner according to 
  system~$\electionsystem$?

\end{description}

The deleting-voter control problems are the following.

\begin{description}
\item[Name:] 
The 
constructive control by deleting voters problem
for $\cale$ elections and the 
destructive control by deleting voters problem
for $\cale$ elections.
\item[Given:] A set $C$ of candidates, a 
collection
$V$ of voters 
  (each voter's vote is a tie-free linear order), 
a
  distinguished candidate $p \in C$, and a nonnegative integer $k$.
\item[Question (constructive):] Is it possible
to by deleting at most $k$ voters ensure that $p$ is a winner of the
resulting $\electionsystem$ election?

\item[Question (destructive):] Is it possible
to by deleting at most $k$ voters ensure that $p$ is not a winner of
the resulting $\electionsystem$ election?
\end{description}

As mentioned in Section~\ref{s:defs}, each partition problem has two
variants, based on how ties are handled in the preliminary (i.e.,
first) round.  In the ties-eliminate tie-handling model, if a
first-round election does not have a unique winner then no one from
that subelection moves forward to the second round.  In the ties-promote
tie-handling model, if a first-round election does not have a unique
winner then all winners (if any) from that subelection move forward to
the second round.  This issue of which tie-handling rule is used is
what is referred to in the definition below (and later, in the 
definitions of 
candidate partitioning) by the phrase ``that survive the
tie-handling rule.''

\begin{description}
\item[Name:] 
The 
constructive control by partition of voters problem
for $\cale$ elections and the 
destructive control by partition of voters problem
for $\cale$ elections.
\item[Given:] A set $C$ of candidates, a 
collection
$V$ of voters
  (each voter's vote is a tie-free linear order), 
and a distinguished candidate
  $p \in C$.
\item[Question (constructive):] Is there
  a partition of $V$ into $V_1$ and $V_2$ such that $p$ is a winner of
  the two-stage election where the winners of election $(C,V_1)$ that
  survive the tie-handling rule compete against the winners of
  $(C,V_2)$ that survive the tie-handling rule?  Each subelection (in
  both stages) is conducted using election system~$\electionsystem$.

\item[Question (destructive):] Is there a
  partition of $V$ into $V_1$ and $V_2$ such that $p$ is not a winner
  of the two-stage election where the winners of election $(C,V_1)$
  that survive the tie-handling rule compete against the winners of
  $(C,V_2)$ that survive the tie-handling rule?  Each subelection (in
  both stages) is conducted using election system~$\electionsystem$.
\end{description}

\subsection{Candidate-Control Definitions}

The unlimited-adding candidate control problems are the following.

\begin{description}
\item[Name:] 
The 
constructive control by adding an unlimited number of candidates problem
for $\cale$ elections and the 
destructive control by adding an unlimited number of candidates problem
for $\cale$ elections.
\item[Given:] Disjoint sets $C$ and $D$ of candidates, a 
collection
$V$
  of voters (each voter's vote is a tie-free linear order over 
the candidates in the set $C \cup D$), and a distinguished candidate
  $p \in C$.
\item[Question (constructive):] Is there
  a subset $E$ of $D$ such that $p$ is a winner of
  the $\electionsystem$ election with voters $V$ and candidates $C
  \cup E$?
\item[Question (destructive):] Is there
  a subset $E$ of $D$ such that $p$ is not a winner
  of the $\electionsystem$ election with voters $V$ and candidates $C
  \cup E$?
\end{description}

The adding-candidate control problems are the following.

\begin{description}
\item[Name:] 
The 
constructive control by adding candidates problem
for $\cale$ elections and the 
destructive control by adding candidates problem
for $\cale$ elections.
\item[Given:] Disjoint sets $C$ and $D$ of candidates, a 
collection
$V$
  of voters (each voter's vote is a tie-free linear order over 
the candidates in the set $C \cup D$), and a distinguished candidate
  $p \in C$, and a nonnegative integer $k$.
\item[Question (constructive):] Is there
  a subset $E$ of $D$ such that $\|E\| \leq k$ and
  $p$ is a winner of the $\electionsystem$ election with voters $V$
  and candidates $C \cup E$?
\item[Question (destructive):] Is there
  a subset $E$ of $D$ such that $\|E\| \leq k$ and
  $p$ is not a winner of the $\electionsystem$ election with voters
  $V$ and candidates $C \cup E$?
\end{description}

The deleting-candidate control problems are the following.

\begin{description}
\item[Name:] 
The 
constructive control by deleting candidates problem
for $\cale$ elections and the 
destructive control by deleting candidates problem
for $\cale$ elections.
\item[Given:] A set $C$ of candidates, a 
collection
$V$ of voters 
  (each voter's vote is a tie-free linear order), a
  distinguished candidate $p \in C$, and a nonnegative integer $k$.
\item[Question (constructive):] Is it possible
to by deleting at most $k$ candidates ensure that $p$ is a winner of
the resulting $\electionsystem$ election?

\item[Question (destructive):] Is it possible
to by deleting at most $k$ candidates other than $p$ ensure that $p$
is not a winner of the resulting $\electionsystem$ election?
\end{description}

The runoff-partition candidate 
control problems are the following.  In runoff-partition candidate
control problems,
each candidate must participate in precisely one of the two first-round 
preliminary elections.
\begin{description}
\item[Name:] 
The 
constructive control by runoff partition of candidates problem
for $\cale$ elections and the 
destructive control by runoff partition of candidates problem
for $\cale$ elections.
\item[Given:] A set $C$ of candidates, a 
collection
$V$ of voters
  (each voter's vote is a tie-free linear order), 
and a distinguished candidate
  $p \in C$.
\item[Question (constructive):] Is there
  a partition of $C$ into $C_1$ and $C_2$ such that $p$ is a winner of
  the two-stage election where the winners of subelection $(C_1,V)$
  that survive the tie-handling rule compete against the winners of
  subelection $(C_2,V)$ that survive the tie-handling rule? Each
  subelection (in both stages) is conducted using election
  system~$\electionsystem$.

\item[Question (destructive):] Is there
  a partition of $C$ into $C_1$ and $C_2$ such that $p$ is not a winner
  of the two-stage election where the winners of subelection $(C_1,V)$
  that survive the tie-handling rule compete against the winners of
  subelection $(C_2,V)$ that survive the tie-handling rule? Each
  subelection (in both stages) is conducted using election
  system~$\electionsystem$.
\end{description}

The partition candidate 
control problems are the following.  In partition candidate
control problems, the partition is between candidates who participate 
in a first-round preliminary election and candidates who get a 
``bye'' through the first round.

\begin{description}
\item[Name:] 
The 
constructive control by partition of candidates problem
for $\cale$ elections and the 
destructive control by partition of candidates problem
for $\cale$ elections.
\item[Given:] A set $C$ of candidates, a
collection
$V$ of voters
  (each voter's vote is a tie-free linear order), 
and a distinguished candidate
  $p \in C$.
\item[Question (constructive):] Is there
  a partition of $C$ into $C_1$ and $C_2$ such that $p$ is a winner of
  the two-stage election where the winners of subelection $(C_1,V)$
  that survive the tie-handling rule compete against all candidates in
  $C_2$? Each subelection (in both stages) is conducted using election
  system~$\electionsystem$.

\item[Question (destructive):] Is there a
  partition of $C$ into $C_1$ and $C_2$ such that $p$ is not a winner
  of the two-stage election where the winners of subelection $(C_1,V)$
  that survive the tie-handling rule compete against all candidates in
  $C_2$? Each subelection (in both stages) is conducted using election
  system~$\electionsystem$.

\end{description}

\end{document}